\documentclass[11pt,onecolumn,draftcls,dvips]{IEEEtran}
\usepackage{amsmath,amssymb,epsfig,color}
\usepackage[breaklinks=true, colorlinks=true, linkcolor=dblue, urlcolor=dblue,
citecolor=dblue, pdfpagemode=None, pdfstartview=FitH]{hyperref}

\newtheorem{theorem}{Theorem}
\newtheorem{lemma}{Lemma}
\definecolor{gray}{cmyk}{.2,0.2,.3,.1}
\definecolor{dred}{cmyk}{0,0.9,0.4,0.3}
\definecolor{dblue}{rgb}{0,0,0.5}
\definecolor{dgreen}{rgb}{0,0.3,0}
\definecolor{dgray}{rgb}{0.3,0.3,0}

\title{On the Scalability of Cooperative Time Synchronization
  in Pulse-Connected Networks
  \thanks{The authors are with the School of Electrical and Computer
  Engineering, Cornell University, Ithaca, NY.  URL:
  \href{http://cn.ece.cornell.edu}{{\tt http://cn.ece.cornell.edu/}}.
  Work supported by the National
  Science Foundation, under awards CCR-0238271 (CAREER), CCR-0330059,
  and ANR-0325556.}}
\author{An-swol Hu \hspace{2cm} Sergio D.\ Servetto}
\date{February 12, 2006.}

\begin{document}
\maketitle

\begin{picture}(0,0)
\put(0,78){\tt\small To appear in the IEEE Transactions on Information
  Theory.}
\end{picture}

\vspace{-10mm}
\begin{abstract}
\it\noindent The problem of time synchronization in dense wireless
networks is considered.  Well established synchronization
techniques suffer from an inherent scalability problem in that
synchronization errors grow with an increasing number of hops
across the network. In this work, a model for communication in
wireless networks is first developed, and then the model is used
to define a new time synchronization mechanism. A salient feature
of the proposed method is that, in the regime of asymptotically
dense networks, it can average out all random errors and maintain
global synchronization in the sense that all nodes in the
multi-hop network can see identical timing signals. This is
irrespective of the distance separating any two nodes.
\end{abstract}

{\keywords Cooperation in networks, large network asymptotics,
relay networks, scalability, sensor networks, time
synchronization, wireless communications.}

\section{Introduction}

\subsection{Time Synchronization in Large Distributed Systems}

The problem of time synchronization in large distributed systems
consists of giving all the physically disjoint elements of the
system a common time scale on which to operate.  This common time
scale is usually achieved by periodically synchronizing the clock
at each element to a reference time source, so that the local time
seen by each element of the system is approximately the same. Time
synchronization plays an important role in many systems in that it
allows the entire system to cooperate and function as a cohesive
group.

Time synchronization is an old problem~\cite{Lamport:78}, but the
question of scalability is not.  Recent advances in sensor
networks show a clear trend towards the development of large scale
networks with high node density.  For example, a hardware
simulation-and-deployment platform for wireless sensor networks
capable of simulating networks with on the order of 100,000 nodes
was recently developed~\cite{KellyEM:03}. As well, for many years
the Smart Dust project sought to build cubic-millimeter motes for
a wide range of applications~\cite{WarnekLLP:01}.  Also, there is
work in progress on the drastic miniaturization of power
sources~\cite{LiLBH:02}. These developments (and many others)
indicate that large scale, high density networks are on the
horizon.

Large scale, high density networks have applications in a variety
of situations.  Consider, for example, the military application of
sniper localization. Large numbers of wireless nodes can be
deployed to find the shooter location as well as the trajectory of
the projectile~\cite{LedecziVMSBNK:05}.  Since the effective range
of a long-range sniper rifle can be nearly $2$km, in order to
fully track the trajectory of the projectile it may be essential
that our deployed network be tightly synchronized over distances
of a few kilometers.  Another example might be the implementation
of a distributed radio for communication.  In extracting
information from a deployed sensor network, it may be beneficial
for the nodes to cooperatively transmit information to a far away
receiver~\cite{BarriacMM:04,OchiaiMPT:05,HuS:03c}.  Such an
application would require that nodes across the network be well
synchronized. As a result, a need for the synchronization of large
distributed systems is very real and one that requires careful
study to understand the fundamental performance limits on
synchronization.

\subsection{Approaches to Synchronization and the Limitations}

The synchronization of large networks has been studied in fields
ranging from biology to electrical engineering.  The study of
synchronous behavior has generally taken one of two approaches.
The first approach is to consider synchronization as an emergent
behavior in complex networks of oscillators. In that work, models
are developed to describe natural phenomena and synchronization
emerges from these models.  The second approach is to develop and
analyze algorithms that synchronize engineering networks.  Nodes
are programmed with algorithms that estimate clock skew and clock
offset to achieve network synchronization.  However, both of these
approaches have significant limitations.

\subsubsection{The Emergence of Synchronous Behavior}

Emergent synchronization properties in large populations has been
the object of intense study in the applied
mathematics~(\cite{MatharM:96,VanVreeswijkA:93}),
physics~(\cite{Chen:94,CorralPDA:95,DiazGuerraPA:98,ErnstPG:98,Gerstner:96,
GuardiolaDLP:00,HerzH:95,Kuramoto:91}), and neural
networks~(\cite{Izhikevich:99,SmithCN:94}) literature. These
studies were motivated by a number of examples observed in nature:
\begin{itemize}
\item In certain parts of south-east Asia, thousands of male fireflies
  congregate in trees and flash in synchrony at night~\cite{BuckB:76}.
\item Pacemaker cells of the heart, which on average cause 80
contractions a minute during a person's lifetime~\cite{Jalife:84}.
\item The insulin-secreting cells of the
pancreas~\cite{ShermanRK:88}.
\end{itemize}
For further information and examples, see~\cite{MirolloS:90,
Strogatz:03,McClintock:71,Walker:69}, and the references therein.

A number of models have been proposed to explain the emergence of
synchrony, but perhaps one of the most successful and well known
is the model of {\em pulse-coupled oscillators} by Mirollo and
Strogatz~\cite{MirolloS:90}, based on dynamical systems theory.
Consider a function $f:[0,1]\to[0,1]$ that is smooth, monotone
increasing, concave down (i.e., $f' > 0$ and $f'' < 0$), and is
such that $f(0)=0$ and $f(1)=1$.  Consider also a phase variable
$\phi$ such that $\partial\phi/\partial t = \frac 1 T$, where $T$
is the period of a cycle.  Then, each element in a group of $N$
oscillators is described by a state variable $x_i\in[0,1]$ and a
phase variable $\phi_i\in[0,1]$ as follows:
\begin{itemize}
\item In isolation, $x_i(t)=f(\phi_i(t))$.
\item If $\phi_i(t) = 0$ then $x_i(t) = 0$, and if $\phi_i(t) = 1$ then
  $x_i(t) = 1$.
\item When $x_i(t_0) = 1$ for any of the $i$'s and some time
$t_0$,
  then for all other $1\leq j\leq N$, $j\ne i$
  \[ \phi_j(t_0^+) = \left\{\begin{array}{rl}
                     f^{-1}(x_j(\phi_j(t_0))+\varepsilon_{i}),
                        & x_j(\phi_j(t_0)) + \varepsilon_{i} \leq 1 \\
                     1, & x_j(\phi_j(t_0)) + \varepsilon_{i} > 1,
                     \end{array}\right.
  \]
  where $t_0^+$ denotes an infinitesimal amount of time after $t_0$.
  That is, oscillator $i$ reaching the end of a cycle causes the state
  of all other oscillators to increase by the amount $\varepsilon_{i}$,
  and the phase variable to change accordingly.
\end{itemize}
The state variable $x_i$ can be thought of as a voltage. Charge is
accumulated over time according to the nonlinearity $f$ and it
discharges once it reaches full charge, resetting the charging
process. Upon discharging, it causes all other charges to increase
by a fixed amount of $\varepsilon_{i}$, up to the discharge point.
For this model, it is proved in~\cite{MirolloS:90} that for all
$N$ and for almost all initial conditions, the system eventually
becomes synchronized.

For the network to converge into a synchronous state, one key
assumption is that the behavior of every single oscillator is
governed by the same function $f(\cdot)$. This means that all
oscillators must have the same frequency. From the literature, it
appears that this requirement is nearly always needed. As far as
we are aware, for a fully synchronous behavior to emerge, the
oscillators need to have the same, or nearly the same, oscillation
frequencies.

The need for nearly identical oscillators presents a significant
limitation for emergent synchronization.  This emergence of
synchrony is clearly desirable and it has been considered for
communication and sensor networks
in~\cite{HongS:04,Werner-AllenTPWN:05, LucarelliW:04}. However,
whether or not these techniques can be adapted to synchronize
networks with nodes that have arbitrary oscillator frequencies
(clock skew) is still unclear.  Thus, in order to overcome this
limitation and find techniques capable of synchronizing a more
general class of networks, we turn to algorithms designed to
estimate certain unknown parameters such as clock skew.

\subsubsection{Estimation of Synchronization Parameters and the Scalability Problem}
\label{sec:estimate-params}

There have been many synchronization techniques proposed for use
in sensor networks.  These algorithms generally allow each node to
estimate its clock skew and clock offset relative to the reference
clock.  Reference Broadcast Synchronization
(RBS)~\cite{ElsonGE:02} eliminates transmitter side uncertainties
by having a transmitter broadcast reference packets to the
surrounding nodes. The receiving nodes then synchronize to each
other using the arrival of the reference packets as
synchronization events. Tiny-Sync/Mini-Sync~\cite{SichitiuV:03}
and the Timing-sync Protocol for Sensor Networks
(TPSN)~\cite{GaneriwalKW:03} organize the network into a
hierarchial structure and the nodes are synchronized using
pair-wise synchronization.  In lightweight tree-based
synchronization (LTS)~\cite{GreunenR:03}, pair-wise
synchronization is also employed but the goal of LTS is to reduce
communication and computation requirements by taking advantage of
relaxed accuracy constraints.  The Flooding Time Synchronization
Protocol (FTSP)~\cite{MarotiKSL:04} achieves one-hop
synchronization by having a root node broadcast timing information
to surrounding nodes.  These surrounding nodes then proceed to
broadcast their synchronized timing information to nodes beyond
the broadcast domain of the root node.  This process can continue
for multi-hop networks.

The problem with each of these traditional synchronization
techniques is that synchronization error will increase with each
hop.  Since each node is estimating certain synchronization
parameters, i.e. clock skew, there will be inherent errors in the
estimate.  As a result, a node multiple hops away from the node
with the reference clock will be estimating its parameters from
intermediate nodes that already have estimation errors. Therefore,
this introduces a fundamental \emph{scalability problem}: as the
number of hops across the network grows, the synchronization error
across the network will also grow.

Current trends in network technology are clearly moving us in the
direction of large, multi-hop networks.  First, sensors are
decreasing in size and this size decrease will most likely be
accompanied by a decrease in communication range.  Thus, more hops
will be required to traverse a network deployed over a given area.
Second, as we deploy networks over larger and larger areas, then
for a given communication range, the number hops across the
network will also increase.  In either case, the increased number
of hops required to communicate across the network will increase
synchronization error.  Therefore, it is essential that we develop
techniques than can mitigate the accumulation of synchronization
error over multiple hops.

\subsection{Spatial Averaging and Synchronization}

\subsubsection{Cooperation through Spatial Averaging}

To decrease the error increase in each hop, we need to decrease
the estimation error.  There are two primary ways of achieving
this.  First, each node can increase the amount of timing
information it obtains from neighboring nodes.  For example, from
a received timing packet, the node may be able to construct a data
point telling it the approximate time at the reference clock and
the corresponding time at its local clock.  Using a collection of
these data points, the node can estimate clock skew and clock
offset.  So instead of using, say, five packets with timing
information, a node can wait for ten packets.  More data points
will generally give better estimates.  The drawback to such an
approach is the increase in the number of packet exchanges.

The second way in which to reduce estimation error is to increase
the quality of each data point obtained by the nodes.  This can be
achieved through improving packet exchange algorithms and time
stamping techniques.  However, we believe that there is one
fundamentally new approach to improving data point quality that
has not be carefully studied.  This is to use spatial averaging to
improve the quality of each data point.

The motivation for this approach is very simple.  Assume that each
node has many neighbors.  If all nodes in the network are to be
synchronized, then the neighbors of any given node will also have
synchronization information.  Is it possible to simultaneously use
information from all the neighbors to improve the quality of a
timing observation made by a node?  Furthermore, it would seem to
make sense that with more neighbors, hence more available timing
information, the quality of the constructed data point should
improve.  If this is indeed the case, then achieving
synchronization through the use of spatial averaging will provide
a fundamentally new trade-off in improving synchronization
performance.  Network designers would simply be able to increase
the number and density of nodes to obtain better network
synchronization.  The study of cooperative time synchronization
using spatial averaging is the focus of this work.

\subsubsection{Model and Technique} \label{sec:introModel}

To obtain a model for developing cooperative synchronization in
large wireless networks, we begin by looking at the signals
observed by a node in a network with $N$ nodes uniformly deployed
over a fixed finite area. To start, we assume propagation delay to
negligible (the general case is considered in
Section~\ref{sec:timeDelay}). All nodes transmit a pulse $p(t)$
and a node $j$ will see a signal $A_{j,N}(t)$ which is the
superposition of all these pulses,
\[
  A_{j,N}(t)
    = \sum_{i=1}^N \frac{A_{max}K_{j,i}}{N}
                   p(t-\tau_{0}-T_{i}).
\]
In this expression, $p(t)$ is the basic pulse transmitted by each
node (assumed to be the same for all nodes).  $\tau_{0}$ is the
ideal pulse transmit time, but since we assume imperfect time
synchronization among the nodes we have $T_{i}$ modelling random
errors in the pulse transmission time.  $K_{j,i}$ models the
amplitude loss in the signal transmitted by the $i$th node.
$A_{max}$ is the maximum magnitude transmitted by a node. We scale
each node's transmission by $N$ so that as the network density
grows, the total power radiated does not grow unbounded.  This
model thus describes the received signal seen at a node $j$ for a
network with $N$ nodes and this holds for any $N$.  Increasing $N$
will have two effects: (a) node density will increase since the
network area is fixed and (b) node signal transmission magnitude
will decrease due to the $1/N$ scaling.  Therefore, by increasing
$N$ this model allows us to study the scalability of networks as
node density grows and node size decreases.

Given that these are the signals observed at each node, we ask: is
it possible for $A_{j,N}(t)$ to encode a time synchronization
signal that will enable all nodes in the network to synchronize
their clocks with bounded error, as $N\to\infty$? The answer is
yes, and the key to proving all our results is the law of large
numbers.

\begin{figure}[!ht]
\centerline{\psfig{file=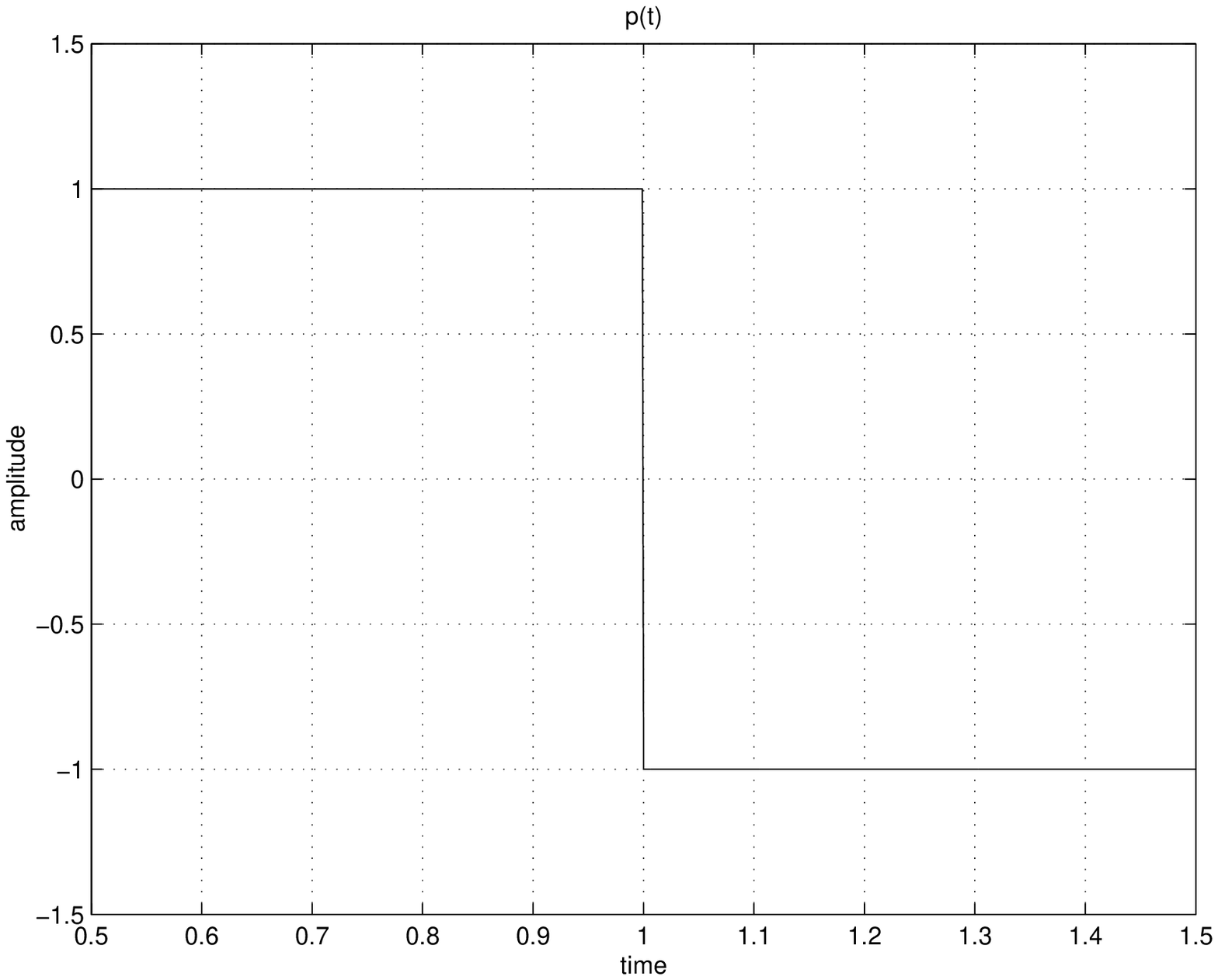,height=4cm,width=7cm}
            \psfig{file=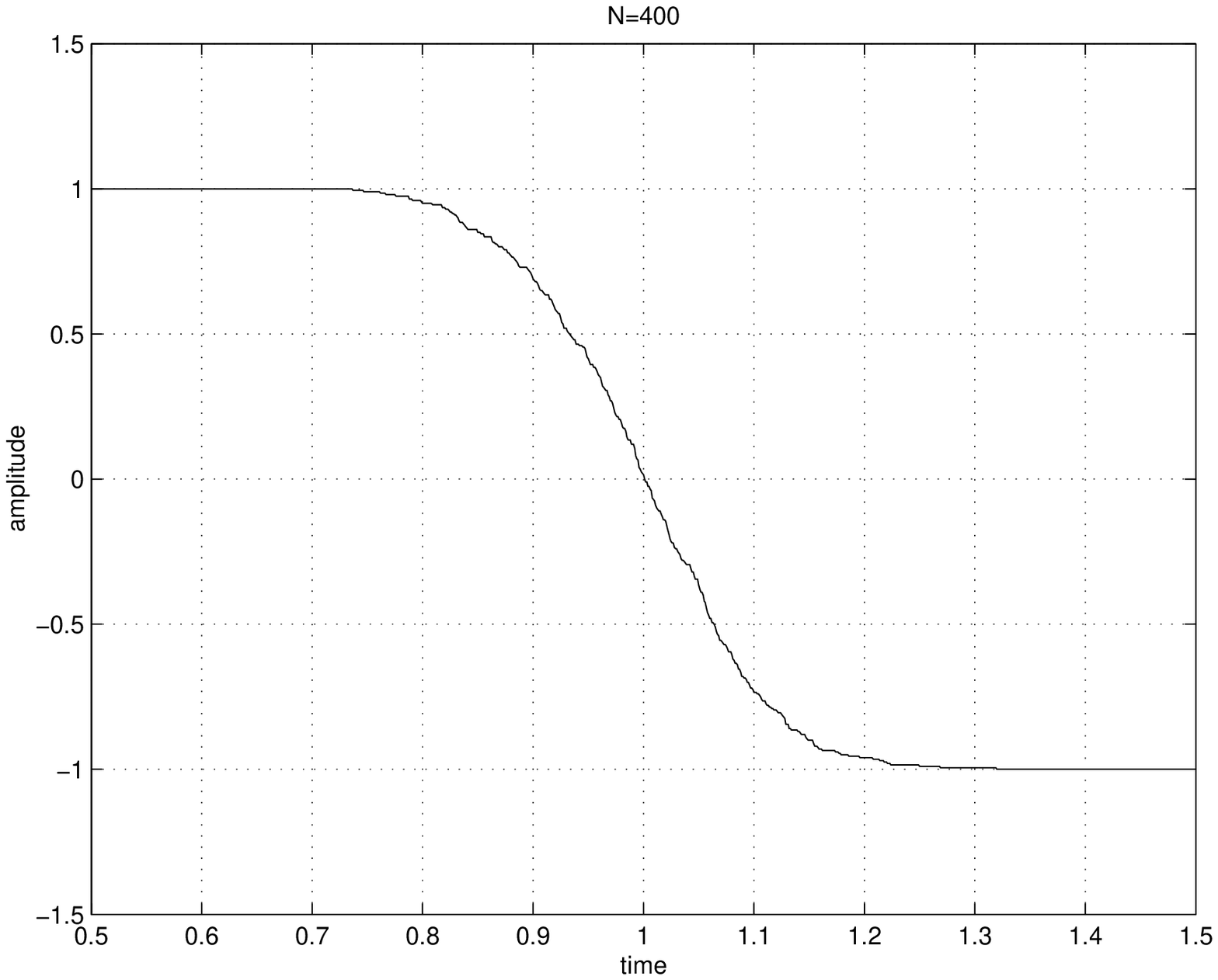,height=4cm,width=7cm}}
\caption{\small Assume $N$ square waves are transmitted (one by
  each node) at random times.  These times have the properties that
  they all have the same mean, a small variance compared to the
  duration of the wave, and their distribution is symmetric.  Then,
  under the assumption of $N$ large, it follows from the Law of
  Large Numbers that the observed signal is going to be a smoothed
  version of the square wave, in which the center {\em
  zero-crossing} will correspond to the location of the mean of the
  random times.}
\label{fig:why-sync-holds}
\end{figure}

Our key idea is the following. If all nodes were able to determine
when time $\tau_{0}$ (in the reference time) arrives, then by
transmitting $p(t)$ at time $\tau_{0}$, the signal observed at any
node $j$ would be $p(t-\tau_{0})\sum_{i=1}^N
\frac{A_{max}K_{j,i}}{N}$, which is a suitably scaled version of
$p(t)$ centered at $\tau_{0}$. In reality however, there will be
some error in the determination of $\tau_{0}$, which we account
for by allowing for a node-dependent random error $T_i$.  But, if
the distribution of $T_i$ satisfies certain conditions, then the
effects of that timing error can be averaged out. A pictorial
representation of why this should be the case is shown in
Fig.~\ref{fig:why-sync-holds}.

Therefore, intuitively we can see how the technique of
\emph{cooperative time synchronization} using spatial averaging
can average out the inherent timing errors in each node. Even
though there is randomness and uncertainty in each node's
estimates, by using cooperation among a large number of nodes it
is possible to recover {\em deterministic} parameters from the
resulting aggregate waveform (such as the location of certain
zero-crossings) in the limit as node density grows unbounded. Thus
more nodes will give us better estimates. This is because the
random waveform converges to a deterministic one as more and more
nodes cooperatively generate an aggregate waveform. At the same
time, the average power required by each node will decrease since
smaller nodes send smaller signals. Therefore, by programming
suitable dynamics into the nodes, in this paper we show how it is
possible to generate an aggregate output signal with {\em
equispaced} zero-crossings in the limit of asymptotically dense
networks. Thus, the detection of these zero-crossings plays the
same role as that of an externally generated time reference signal
based on which all nodes can synchronize.

We develop this synchronization technique in three main steps.
One, we set up the model for $A_{j,N}(t)$.  Two, we specify
characteristics of the model (i.e. the distribution of $T_{i}$)
that allow us to prove desirable properties of the aggregate
waveform (such as a center zero-crossing at $\tau_{0}$).  Three,
we develop the estimators needed for our synchronization technique
and show that the estimators give us the desired characteristics.

\subsection{Main Contributions and Organization of the Paper}

The main contributions presented in this paper are the following;
\begin{itemize}
\item The definition of a probabilistic model for the study of the
time synchronization problem in wireless networks.  This model
does contain the classical Mirollo-Strogatz model as a special
case, but its formulation and the tools used to prove convergence
results are of a completely different nature (purely
probabilistic, instead of based on the theory of dynamical
systems). \item Using this model, we provide a rigorous analysis
of a new cooperative time synchronization technique that employs
spatial averaging and has favorable scaling properties. As the
density of nodes increases, synchronization performance improves.
In particular, in the limit of infinite density, deterministic
parameters for synchronization can be recovered. \item We show
that cooperative time synchronization works perfectly for
negligible propagation delay. When propagation delay is
considered, we find that asymmetries at the boundaries reveal some
limitations that need to be carefully considered in designing
algorithms that take advantage of spatial averaging.
\end{itemize} In analyzing the
proposed cooperative time synchronization technique, our goal is
to show that the proposed technique can average out all random
error and provide deterministic parameters for synchronization as
node density grows unbounded. This asymptotic result can be viewed
as a \emph{convergence in scale} to synchrony.  The result serves
as a theoretical foundation for allowing a new trade-off between
node density and synchronization performance.  In particular,
higher node density can yield better synchronization.

The rest of this paper is organized as follows.  The general model
is presented in Section~\ref{sec:systemModel}. Of particular
interest here is Section~\ref{sec:specialcase}, where we show how
our model contains the model of Mirollo and Strogatz for
pulse-coupled oscillators as a special case~\cite{MirolloS:90}. In
Section~\ref{sec:timesync-setup} we specialize the general model
for our synchronization setup and develop waveform properties that
will be used in time synchronization. In
Section~\ref{sec:timeSynchronization} we develop the cooperative
time synchronization technique for no propagation delay.  We
extend the cooperative synchronization ideas to the case of
propagation delay in Section~\ref{sec:timeDelay}. The paper
concludes in Section~\ref{sec:conclusion} with a detailed
discussion on the scalability issue and how the technique proposed
in this work lays the theoretical foundation for a general class
of cooperative time synchronization techniques that use spatial
averaging.

\section{System Model}
\label{sec:systemModel}

\subsection{Clock Model}

We consider a network with $N$ nodes uniformly distributed over a
fixed finite area. The behavior of each node $i$ is governed by a
clock $c_{i}$ that counts up from $0$. The introduction of $c_{i}$
is important since it provides a consistent timescale for node
$i$. By maintaining a table of pulse-arrival times, node $i$ can
utilize the arrival times of many pulses over an extended period
of time.

The clock of one particular node in the network will serve as the
reference time and to this clock we wish to synchronize all other
nodes. We will call the node with the reference clock node $1$ and
the clocks of other nodes are defined relative to the clock of
node $1$.  We never adjust the frequency or offset of the local
clock $c_{i}$ because we wish to maintain a consistent time scale
for node $i$.

The clock of node $1$, $c_{1}$, will be defined as $c_{1}(t)=t$
where $t \in [0,\infty)$. Taking $c_{1}$ to be the reference
clock, we now define the clock of any other arbitrary node $i$,
$c_{i}$. We define $c_{i}$ as
\begin{equation} \label{eq:clock}
c_{i}(t)=\alpha_{i}(t-\bar{\Delta}_{i})+\Psi_{i}(t),
\end{equation}
where
\begin{itemize}
\item $\bar{\Delta}_{i}$ is an unknown offset between the start
  times of $c_{i}$ and $c_{1}$.
\item $\alpha_{i}>0$ is a constant and for each $i$,
  $\alpha_{i} \in [\alpha_{low},\alpha_{up}]$ where
  $\alpha_{up},\alpha_{low}>0$ are finite. This bound on $\alpha_{i}$
  means that the frequency offsets between any two nodes can not be
  arbitrarily large.
\item $\Psi_{i}(t)$ is a stochastic process modeling random timing
  jitter.
\end{itemize}
Thus, this model assumes that there is a bounded constant
frequency offset between the oscillators of any two nodes as well
as some random clock jitter.

It is important to note that node $1$ does not have to be special
in any way; its clock is simply a reference time on which to
define the clocks of the other nodes.  This means that our clock
model actually describes the relative relationship of all the
clocks in the network by using an arbitrary node's clock as a
reference.

\subsection{Pathloss Only Model} \label{sec:prop-model}

\subsubsection{A Random Model for Pathloss}

From Section~\ref{sec:introModel}, we see that we are interested
in studying the aggregate waveform observed at a node $j$.  As a
result, we are only concerned with the aggregate signal magnitude
and do not care about the particular signal contribution from each
surrounding node.  With this in mind, we can develop a random
model for pathloss that, for dense networks, gives the appropriate
aggregate signal magnitude at node $j$. Such a model is ideal for
our situation since we are studying asymptotically dense networks.

We start with a general pathloss model $K(d)$, where $0\leq
K(d)\leq 1$ for all distances $d\geq 0$, is non-increasing and
continuous. $K(d)$ is a fraction of the transmitted magnitude seen
at distance $d$ from the transmitter. For example, if the receiver
node $j$ is at distance $d$ from node $i$, and node $i$ transmits
a signal of magnitude $A$, then node $j$ will hear a signal of
magnitude $AK(d)$. We derive $K(d)$ from a power pathloss model
since any pathloss model captures the \emph{average} received
power at a given distance from the transmitter.  This average
received power is perfect for modelling received signal magnitudes
in our problem setup since we are considering asymptotically dense
networks. Due to the large number of nodes at any given distance
$d$ from the receiver, using the average received magnitude at
distance $d$ as the contribution from each node at that distance
will give a good modelling of the amplitude of the aggregate
waveform.

The random pathloss variable $K_{j}$ will be derived from $K(d)$.
To understand how $K_{j}$ and $K(d)$ are related, we give an
intuitive explanation of the meaning of $K_{j}$ as follows: the
$\textrm{Pr}[K_{j}\in(k,k+\Delta)]$ is the fraction of nodes at
distances $d$ from node $j$ such that $K(d)\in(k,k+\Delta)$, where
$\Delta$ is a small constant. This means that, roughly speaking,
for any given scaling factor $K_{j}=k$, $f_{K_{j}}(k)\Delta$ is
the fraction of received signals with magnitude scaled by
approximately $k$, where $f_{K_{j}}(k)$ is the probability density
function of $K_{j}$. Thus, if we scale the transmit magnitude $A$
from every node $i$ by an independent $K_{j}$, then as the number
of nodes, $N$, gets large, node $j$ will see $Nf_{K_{j}}(k)\Delta$
signals of approximate magnitude $Ak$, and this holds for all $k$
in the range of $K_{j}$. This is because taking a large number of
independent samples from a distribution results in a good
approximation of the distribution.

Thus, this intuition tells us that scaling the magnitude of the
signal transmitted from every node $i$ by an independent sample of
the random variable $K_{j}$ gives an aggregate signal at node $j$
that is the same magnitude as if we generated the signal using
$K(d)$ directly. Even though the signals from two nodes at the
same distance from a receiver have correlated magnitudes, we do
not care about the signal magnitude from any particular node but
only that an appropriate number of all possible received signal
magnitudes contribute to the aggregate waveform. For a receiving
node $j$, we choose therefore to work with the random variable
$K_{j}$ instead of directly with $K(d)$ because, for the goals of
this paper, doing so has two major advantages: (a) we can obtain
desirable limit results by placing very minimal restrictions on
the distribution of the $K_{j}$'s (and hence on $K(d)$) and (b) we
can apply tools from probability theory (basically, the strong law
of large numbers) to carry out our analysis.

\subsubsection{Definition of $K_{j}$}

\begin{figure}[!h]
\centerline{\psfig{file=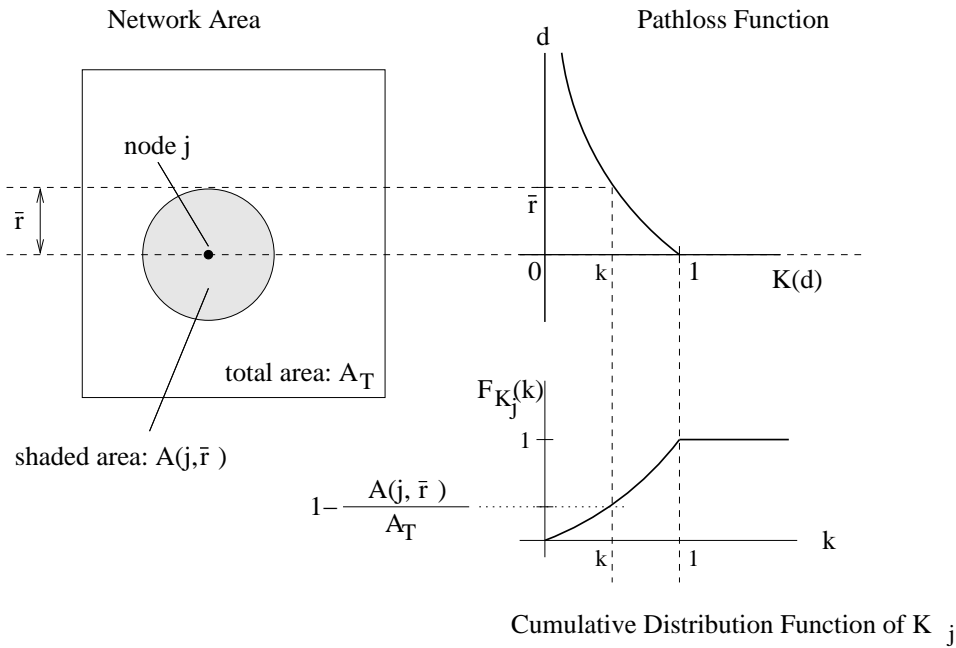,height=8.5cm}}
\caption[An illustration of node $j$, $K(d)$, $\bar{r}$, and
$F_{K_{j}}(k)$.]{\small An illustration of the cumulative
distribution function $F_{K_{j}}(k)$ is shown in the bottom-right
figure.  For a given scaling value $k$, $F_{K_{j}}(k)$ is defined
to be $1-(A(j,\bar{r})/A_{T})$, where the relationship between
$\bar{r}$ and $k$ is shown in the top-right figure. The area
$A(j,\bar{r})$ and its relation to node $j$ is shown in the
top-left figure.} \label{fig:propModel-illus}
\end{figure}

From the above intuition we can define the cumulative distribution
function of $K_{j}$ as
\begin{equation} \label{eq:prop_cdf}
F_{K_{j}}(k) \;\;=\;\; \textrm{Pr}(K_{j}\leq k)
   \;\;=\;\;  \left\{ \begin{array}{ll}
0 & k\in(-\infty,0) \\
   \frac{A_{T}-A(j,\bar{r})}{A_{T}}
   \;\;=\;\; 1 - \frac{A(j,\bar{r})}{A_{T}} & k\in[0,1] \\
   1 & k\in(1,\infty)
   \end{array} \right.
\end{equation}
where
\begin{itemize}
\item $A_{T}$ is the total area of the network, \item $A(j,a)$ is
the area of the network contained in a circle of radius $a$
centered at node $j$, \item $\bar{r} = \textrm{sup}\{d:K(d) >
k\}$.
\end{itemize}
From the above discussion we see that the distribution of $K_{j}$
is only a function of node $j$, the receiving node. We illustrate
the relationship among node $j$, $K(d)$, $\bar{r}$, and
$F_{K_{j}}(k)$ in Fig~\ref{fig:propModel-illus}. We sometimes
write $K_{j,i}$ with $i$ used to index each node surrounding node
$j$. $i$ is thus indexing a sequence of independent random
variables $K_{j,i}$ for fixed $j$. Therefore, for a given $j$,
$K_{j,i}$'s are independent and identically distributed (i.i.d.)
with a cumulative distribution function given by
(\ref{eq:prop_cdf}) for all $i$.

We assume that $K_{j}$ has the following properties:
\begin{itemize}
\item $K_{j}$ is independent from $\Psi_{l}(t)$ for all $j$, $l$,
and $t$. \item $0 \leq K_{j} \leq 1$, $0<E(K_{j})\leq 1$, and
$\textrm{Var}(K_{j})\leq 1$.
\end{itemize}
The requirements on the random variable $K_{j}$ places
restrictions on the model $K(d)$. Any function $K(d)$ that yields
a $K_{j}$ with the above requirements can be used to model
pathloss.

\subsection{Delay and Pathloss Model} \label{sec:delay-model}

In this section we develop a more complex model to simultaneously
model propagation delay and pathloss.  This leads to the joint
development of the delay random variable $D_{j}$ and a
corresponding pathloss random variable $K_{j}$.

\subsubsection{Correlation Between Delay and Pathloss}

Since we want to develop a model for both pathloss and time delay,
we start by keeping the pathloss function $K(d)$ defined in
Section~\ref{sec:prop-model}.  The general delay model assumes a
function $\delta(d)$ that models the time delay as a function of
distance. $\delta(d)$ describes the time in terms of $c_{1}$ that
it takes for a signal to propagate a distance $d$. For example, if
node $i$ and node $j$ are distance $d_{0}$ apart, then a pulse
sent by node $i$ at time $c_{1}=0$ will be seen at node $j$ at
time $c_{1}=\delta(d_{0})$. We make the reasonable assumption that
$\delta(d)$ is continuous and strictly monotonically increasing
for $d\geq 0$.

As with the pathloss only model, we want to define a delay random
variable $D_{j}$ for each receiving node $j$.  Recall that this
means that for every node $j$ there is a random variable $D_{j}$
associated with it since, in general, each node $j$ will see
different delays.  There is a correlation between the delay random
variable $D_{j}$ and the pathloss random variable $K_{j}$. This
correlation arises for two main reasons. First, since in
Section~\ref{sec:prop-model} we define $K(d)$ to be monotonically
decreasing and continuous, it is possible for $K(d)=0$ for
$d\in[R,\infty)$, $R>0$. This might be the case for a multi-hop
network. In this situation, there will be a set of nodes whose
transmissions will never reach node $j$ (i.e. infinite delay) even
though according to $\delta(d)$ these nodes should contribute a
signal with finite delay.  Second, a small $K_{j}$ value would
represent a signal from a far away node. As a result, the
corresponding $D_{j}$ value should be large to reflect large
delay.  Therefore, keeping these two points in mind, we proceed to
develop a model for both pathloss and propagation delay.

\subsubsection{Definition of $D_{j}$ and $K_{j}$}

We define the cumulative distribution function of $D_{j}$ as
\begin{equation} \label{eq:delay_cdf}
F_{D_{j}}(x) \;\;=\;\; \textrm{Pr}(D_{j}\leq x)
    \;\;=\;\; \left\{ \begin{array}{ll}
    0 & x\in(-\infty,0) \\
    \frac{A(j,r')}{A_T} & x\in[0,\delta(R)] \\
    a(x-\delta(R))+\frac{A(j,R)}{A_{T}} & x\in(\delta(R),\delta(R+\Delta
    R)] \\
    1 & x\in(\delta(R+\Delta R),\infty)
    \end{array} \right.
\end{equation}
where $r' = \textrm{sup}\{r:\delta(r)\leq x\}$, $\Delta R>0$ is a
constant, $R=\sup\{d:K(d)>0\}$, and
\begin{displaymath}
a = \frac{1-\frac{A(j,R)}{A_{T}}}{\delta(R+\Delta R)-\delta(R)}.
\end{displaymath}
Recall that $A(j,a)$, defined in Section~\ref{sec:prop-model}, is
the area of the network contained in a circle of radius $a$
centered at node $j$ and $A_{T}$ is the total area of the network.
Note that $R$ can be infinite.

Using the delay random variable $D_{j}$ with the cumulative
distribution function in (\ref{eq:delay_cdf}), we define $K_{j}$
as
\begin{equation} \label{eq:DjKjconnection}
K_{j} = K(\delta^{-1}(D_{j})),
\end{equation}
where $K(\cdot)$ is the deterministic pathloss function from
Section~\ref{sec:prop-model} and
$\delta^{-1}:[0,\infty)\to[0,\infty)$ is the inverse function of
the deterministic delay function $\delta(\cdot)$. Note that
$\delta^{-1}(\cdot)$ exists since $\delta(\cdot)$ is continuous
and strictly monotonically increasing on $[0,\infty)$.

\subsubsection{Intuition Behind $D_{j}$ and $K_{j}$}

\begin{figure}[!h]
\centerline{\psfig{file=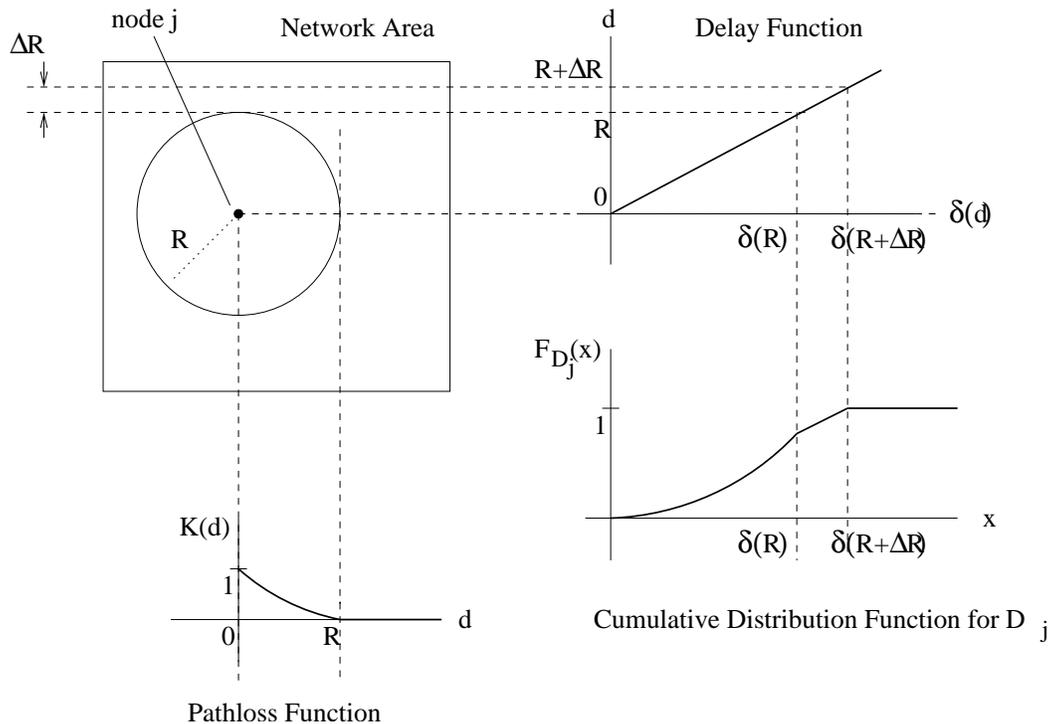,height=9.5cm}}
\caption[An illustration $K(d)$, $\delta(d)$, node $j$, and
$F_{D_{j}}(x)$.]{\small From the top-left and bottom-left figures,
we can see how $K(d)$ determines the set of nodes surrounding node
$j$ that will contribute to the aggregate waveform at node $j$.
This contributing set of nodes is related to $F_{D_{j}}(x)$
through $\delta(d)$ and this is illustrated in the top-right and
bottom-right figures.} \label{fig:delayModel-illus}
\end{figure}

To understand the distribution of $D_{j}$, we need to consider the
definition of $K_{j}$ as well.  Recall that a signal arriving with
delay $D_{j}$ is scaled by the pathloss random variable $K_{j}$.
Let us consider the cumulative distribution in two pieces,
$x\in[0,\delta(R)]$ and $x\in(\delta(R),\infty)$. The case for
$x\in(-\infty,0)$ is trivial.  First, for $x\in[0,\delta(R)]$, the
probability that $D_{j}$ takes a value less than or equal to $x$
is simply the fraction of the network area around node $j$ such
that the nodes are at distances $d$ with $\delta(d)\leq x$. The
intuition is the same as that for the development of $K_{j}$ in
Section~\ref{sec:prop-model}. Second, for
$x\in(\delta(R),\infty)$, the situation is more complex. Note that
a transmitted signal from a node at distance $d\in(R,\infty)$ from
$j$ will arrive at node $j$ with infinite delay since $K(d)=0$ for
$d\in(R,\infty)$. Since any delay values in
$x\in(\delta(R),\infty)$ correspond to distances
$d=\delta^{-1}(x)\in(R,\infty)$, the corresponding scaling value
will be zero because $K_{j}$ and $D_{j}$ are related by
(\ref{eq:DjKjconnection}). As a result, it does not matter what
delay values we assign to the fraction of the network area outside
a circle of radius $R$ centered at node $j$ as long as their delay
value $x$ is such that $\delta^{-1}(x)\in(R,\infty)$.  Thus, we
can arbitrarily choose a constant $\Delta R$ value and construct a
piecewise linear portion of the cumulative distribution function
of $D_{j}$ on $x\in(\delta(R),\infty)$.  The probability that
$D_{j}\in(\delta(R),\infty)$ will be the fraction of the network
area outside a circle of radius $R$ around node $j$.  And since
$D_{j}\in(\delta(R),\infty)$ will have a corresponding $K_{j}$
value that is zero, this fraction of nodes will not contribute to
the aggregate waveform at node $j$.  It is clear that the
correlated $D_{j}$ and $K_{j}$ random variables work together to
accurately model a signal arriving with both pathloss and
propagation delay.  An illustration of how $K(d)$, $\delta(d)$,
node $j$, and $F_{D_{j}}(x)$ are related can be found in
Fig.~\ref{fig:delayModel-illus}.

We require that $D_{j}$ is bounded, has finite expectation, and
has finite variance for all $j$. Note that $D_{j}\geq 0$ by the
requirement that $\delta(d)\geq 0$.  As well, since the cumulative
distribution in (\ref{eq:delay_cdf}) is continuous, and often
absolutely continuous, we assume that $D_{j}$ has a probability
density function $f_{D_{j}}(x)$.  When we write $D_{j,i}$, the $i$
indexes each node surrounding node $j$. Thus, the $D_{j,i}$'s are
independent and identically distributed in $i$ for a given $j$ and
have a cumulative distribution given by (\ref{eq:delay_cdf}).
Using the $K_{j}$ and $D_{j}$ developed in this section to
simultaneously model pathloss and propagation delay, respectively,
we will be able to closely approximate the received aggregate
waveform at any node $j$ as $N\to\infty$.

To summarize, we see that our choice of the pathloss and delay
random variables will depend on what we want to model. If we only
consider pathloss and not propagation delay, then we will use the
random variable $K_{j}$ defined in Section~\ref{sec:prop-model}.
If we account for both pathloss and delay, then we will use the
delay random variable $D_{j}$ in this section
(Section~\ref{sec:delay-model}) and the pathloss random variable
$K_{j}$ defined by (\ref{eq:DjKjconnection}).

\subsection{Synchronization Pulses and the Pulse-Connection Function}

The exchange of pulses is the method through which the network
will maintain time synchronization. Each node $i$ will
periodically transmit a scaled pulse $A_{i} p(t)$, where $A_{i}$
is a constant and $p(t)$, in general, can be any pulse. We call
the interval of time during which a synchronization pulse is
transmitted a \emph{synchronization phase}.

What each node does with a set of pulse arrival observations is
determined by the pulse-connection function $X_{n,i}^{c_{i}}$ for
node $i$.  The pulse-connection function is a function that
determines the time, in the time scale of $c_{i}$, when node $i$
will send its $n$th pulse. It can be a function of the current
value of $c_{i}(t)$ and past pulse arrival times. This function
basically determines how any node $i$ reacts to the arrival of a
pulse.

\subsection{An Example: Pulse-Coupled Oscillators}
\label{sec:specialcase}

The system model that we presented thus far is powerful because it
is very general. In this section we show that it is a
generalization of the pulse-coupled oscillator model proposed by
Mirollo and Strogatz~\cite{MirolloS:90}. As a result, the results
presented in that paper will hold under the simplified version of
our model.

\subsubsection{Model Parameters for Pulse-Coupled Oscillators}

In setting up the system model, Mirollo and Strogatz make four key
assumptions:
\begin{itemize}
\item Pathloss Model: The first assumption that is made is that
there is all-to-all coupling among all $N$ oscillators. This means
that each oscillator's transmission can be heard by all other
oscillators. Thus, for our model we ignore pathloss, i.e. $K(d) =
1$, to allow any node's transmission to be heard by each of the
other $N-1$ nodes. \item Delay Model: The second assumption is
that there is instantaneous coupling.  This assumption is the same
as setting $\delta(d) = 0$. In such a situation we would use our
pathloss only model. \item Synchronization Pulses: The third key
assumption made in~\cite{MirolloS:90} is that there is non-uniform
coupling, meaning that each of the $N$ oscillators fire with
strengths $\epsilon_{1},\dots,\epsilon_{N}$. We modify the
parameters in our model by making node $i$ transmit with magnitude
$A_{i} = \epsilon_{i}$. They also assume that any two pulses
transmitted at different times will be seen by an oscillator as
two separate pulses.  In our model, we may choose any pulse $p(t)$
that has an arbitrarily short duration and each node will detect
the pulse arrival time and pulse magnitude. \item Clock Model: The
fourth important assumption made by Mirollo and Strogatz is that
the oscillators are identical but they start in arbitrary initial
conditions.  We simplify our clock model in~(\ref{eq:clock}) by
eliminating any timing jitter, i.e. $\Psi_{i}(t) = 0$, and making
the clocks identical by setting $\alpha_{i} = 1$ for
$i=1,\dots,N$. We leave $\bar{\Delta}_{i}$ in the model to account
for the arbitrary initial conditions. We also assume that the
phase variable in the pulse-coupled oscillator model increases at
the same rate as our clock.  That is, the time it takes the phase
variable to go from zero to one and the time it takes our clock to
count from one integer value to the next are the same.
\end{itemize}
Now that we have identical system models, what remains is to
modify our model to mimic the coupling action detailed
in~\cite{MirolloS:90}.  This is accomplished by defining a proper
pulse-connection function $X_{n,i}^{c_{i}}$.

\subsubsection{Choice of Pulse-Connection Function}

To match the coupling action in~\cite{MirolloS:90}, we choose a
pulse transmit time function $X_{n,i}^{c_{i}}(z_{k,i}^{c_{i}},
z_{k-1,i}^{c_{i}},\dots,z_{1,i}^{c_{i}}, x_{n-1,i}^{c_{i}})$ that
is a function of pulse receive times and also the time of node
$i$'s $(n-1)$th pulse transmission time. $z_{k,i}^{c_{i}}$ is the
time in terms of $c_{i}$ that node $i$ receives its $k$th pulse
since its last pulse transmission at $x_{n-1,i}^{c_{i}}$.  In this
case, $X_{n,i}^{c_{i}}$ will be a function that updates node $i$'s
$n$th pulse transmission time each time node $i$ receives a pulse.
Let $X_{n,i}^{c_{i}}(k) \stackrel{\small{\Delta}}{=}
X_{n,i}^{c_{i}}(z_{k,i}^{c_{i}},
z_{k-1,i}^{c_{i}},\dots,z_{1,i}^{c_{i}}, x_{n-1,i}^{c_{i}})$ where
it is node $i$'s $n$th pulse transmission time after observing $k$
pulses since its last pulse transmission.  Node $i$ will transmit
its pulse as soon as $X_{n,i}^{c_{i}}\leq c_{i}(t)$ where
$c_{i}(t)$ is node $i$'s current time. As soon as the node
transmits a pulse at $X_{n,i}^{c_{i}}$ the function will reset and
become $X_{n+1,i}^{c_{i}}(0) = x_{n,i}^{c_{i}}+1$. The node is now
ready to receive pulses and at its first received pulse, the next
transmission time will become $X_{n+1,i}^{c_{i}}(1)$.
$X_{n,i}^{c_{i}}$ will thus be defined as
\begin{eqnarray} \label{eq:pco_estimator}
X_{n,i}^{c_{i}}(k) &=& X_{n,i}^{c_{i}}(k-1) -
[f^{-1}(\epsilon_{j}+f(z_{k,i}^{c_{i}}-x_{n-1,i}^{c_{i}}))-(z_{k,i}^{c_{i}}-x_{n-1,i}^{c_{i}})],
\qquad k>0\\
X_{n,i}^{c_{i}}(0) &=& x_{n-1,i}^{c_{i}}+1
\label{eq:pco_estimator2}
\end{eqnarray}
where the pulse received at $z_{k,i}^{c_{i}}$ is a pulse of
magnitude $\epsilon_{j}$ and the function $f:[0,1]\to [0,1]$ is
the smooth, monotonic increasing, and concave down function
defined in~\cite{MirolloS:90}.

\begin{figure}[!h]
\centerline{\psfig{file=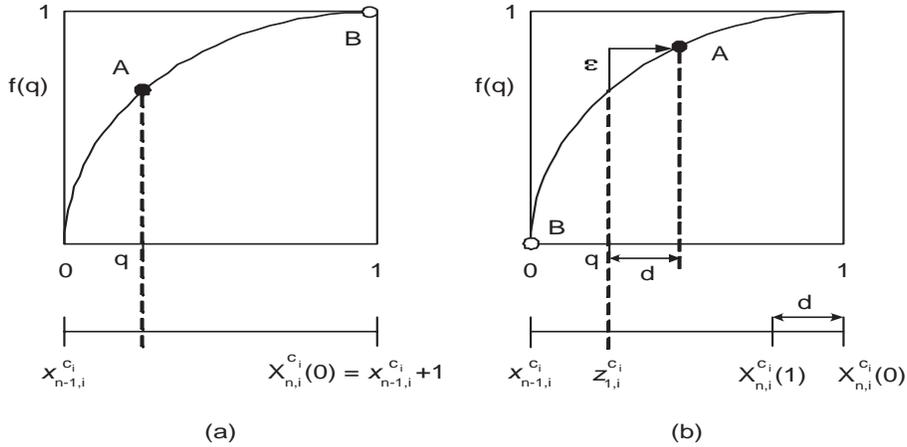,height=6cm,width=12cm}}
\vspace{-5mm} \caption[The connection with pulse-coupled
oscillators.]{\small We illustrate the connection between the
pulse-coupled oscillator coupling model and our clock model.  In
(a), oscillator $B$ is just about to fire and oscillator $A$ has
phase $q$.  In (b), oscillator $B$ fires and increases the phase
of oscillator $A$ by $d$.  This $d$ increase in phase effectively
decreases the time at which $A$ will next fire.  We capture this
time decrease by decreasing the firing time of our node by an
amount $d$. Thus, oscillator $A$ and our node will fire at the
same time.} \label{fig:strogatzModel}
\end{figure}

Equations~(\ref{eq:pco_estimator}) and (\ref{eq:pco_estimator2})
fundamentally say that each time node $i$ receives a pulse, node
$i$'s next transmission time will be adjusted.  This is in line
with the behavior of the coupling model described by Mirollo and
Strogatz since each time an oscillator receives a pulse, its state
variable is pulled up by $\epsilon$ thus adjusting the time at
which the oscillator will next fire. To see how
equations~(\ref{eq:pco_estimator}) and (\ref{eq:pco_estimator2})
relate to the coupling model in~\cite{MirolloS:90}, let us
consider an example with two pulse coupled oscillators. Consider
two oscillators $A$ and $B$ illustrated in
Fig.~\ref{fig:strogatzModel}. In Fig.~\ref{fig:strogatzModel}(a),
we have that oscillator $A$ is at phase $q$ and oscillator $B$ is
just about to fire.  Below the pulse-coupled oscillator model we
have a time axis for node $i$ corresponding to our clock model
going from time $x_{n-1,i}^{c_{i}}$ to $x_{n-1,i}^{c_{i}}+1$. Our
time axis for node $i$ models the behavior of oscillator $A$, that
is, we want node $i$ to behave in the same way as oscillator $A$
under the influence of oscillator $B$. If oscillator $B$ did not
exist, then the phase variable $q$ will match our clock in that
$q$ reaches $1$ at the same time our clock reaches
$X_{n,i}^{c_{i}}(0) = x_{n-1,i}^{c_{i}}+1$ and oscillator $A$ will
fire at the same time our model fires.  In
Fig.~\ref{fig:strogatzModel}(b), oscillator $B$ has fired and has
pulled the state variable of oscillator $A$ up by $\epsilon$. This
coupling has effectively pushed the phase of oscillator $A$ to
$q+d$ and decreased the time before $A$ fires.  In fact, the time
until oscillator $A$ fires again is decreased by $d$.  We can
capture this coupling in our model since we can calculate the lost
time $d$.  The time at which oscillator $B$ fires is
$z_{1,i}^{c_{i}}$ and it is clear that $d =
f^{-1}(\epsilon+f(z_{1,i}^{c_{i}}-x_{n-1,i}^{c_{i}}))-(z_{1,i}^{c_{i}}-x_{n-1,i}^{c_{i}})$.
Thus, if the time that oscillator $A$ will fire again is decreased
by time $d$ due to the pulse of $B$, then we adjust our node
firing time by decreasing the firing time to $X_{n,i}^{c_{i}}(1) =
x_{n-1,i}^{c_{i}}+1-d$.  This is exactly the expression
in~(\ref{eq:pco_estimator}) for $k=1$.  This relationship between
our model for calculating the node firing time and the
pulse-coupled oscillator coupling model can be easily extended to
$N$ oscillators.

We can see then that the pulse-coupled oscillator model
proposed by Mirollo and Strogatz in~\cite{MirolloS:90} is a
special case of our model.  Our model generalizes this
pulse-coupled oscillator model by considering timing jitter,
pulses of finite width, propagation delay, non-identical clocks,
and an ability to accommodate arbitrary coupling functions.

\section{Cooperative Time Synchronization Setup}
\label{sec:timesync-setup}

Just as we could specialize our model to the pulse-coupled
oscillator model of Mirollo and Strogatz, we now specialize the
model for our proposed synchronization technique.  We start under
the assumption of no propagation delay and develop the
synchronization technique for this case. Propagation delay is
considered in Section~\ref{sec:timeDelay}. We proceed in three
steps. In Section~\ref{sec:signal-reception}, we specify the model
for $A^{c_{1}}_{j,N}(t)$, the received waveform at any node $j$.
Second, in Section~\ref{sec:pulseProperties}, we prove that given
certain characteristics of the model, $A^{c_{1}}_{j,N}(t)$ has
very useful limiting properties.  Third, we show in
Section~\ref{sec:timeSynchronization} that estimators (i.e., the
pulse connection function) developed for our synchronization
technique give $A^{c_{1}}_{j,N}(t)$ the desired properties.

\subsection{System Parameters}  \label{sec:systemParameters}

For our synchronization technique, we specialize the general model
by making the following assumptions on $\alpha_{i}$ and
$\Psi_{i}(t)$ for $i=1\dots N$:
\begin{itemize}
\item A characterization of the $\{\alpha_{i}\}$ is given by a
known function $f_{\alpha}(s)$ with
$s\in[\alpha_{low},\alpha_{up}]$ that gives the percentage of
nodes with any given $\alpha$ value. Thus, the fraction of nodes
with $\alpha$ values in the range $s_{0}$ to $s_{1}$ can be found
by integrating $f_{\alpha}(s)$ from $s_{0}$ to $s_{1}$. We assume
that $|f_{\alpha}(s)|<G_{\alpha}$, for some constant $G_{\alpha}$.
We keep this function constant as we increase the number of nodes
in the network ($N\to\infty$). Given any circle of radius $R$ that
intersects the network, the nodes within that circle will have
$\alpha_{i}$'s that are characterized by $f_{\alpha}(s)$. $R$ is
the maximum $d$ such that $K(d)>0$. This means that the set of
nodes that any node $j$ will hear from will have its
$\alpha_{i}$'s characterized by a known function. Note that $R$
can be infinite, and in that case, any node $j$ hears from all
nodes in the network. Fundamentally, $f_{\alpha}(s)$ means that as
we increase node density, the new nodes have $\alpha$ parameters
that are well distributed in a predictable manner. \item
$\Psi_{i}(t)$ is a zero mean Gaussian process with samples
$\Psi_{i}(t_{0})\sim {\mathcal N}(0,\sigma^{2})$, for any $t_{0}$,
and independent and identically distributed samples for any set of
times $[t_{0},\ldots,t_{k}]$, $k$ a positive integer. We assume
$\sigma^{2}<\infty$ and note that $\sigma^{2}$ is defined in terms
of the clock of node $i$. We assume that $\Psi_{i}(t)$ is Gaussian
since the RMS (root mean square) jitter is characterized by the
Gaussian distribution~\cite{Roberts:03}.
\end{itemize}
We maintain the full generality of the pathloss model from
Section~\ref{sec:prop-model}.  Note that throughout this work we
assume no transmission delay or time-stamping error. This means
that a pulse is transmitted at exactly the time the node intends
to transmit it. We make this assumption since there will be no
delay in message construction or access time~\cite{ElsonGE:02}
because our nodes broadcast the same simple pulse without worrying
about collisions. Also, when a node receives a pulse it can
determine its clock reading without delay since any time stamping
error is small and can be absorbed into the random jitter.

\subsection{Signal Reception Model} \label{sec:signal-reception}

For our proposed synchronization technique, the aggregate waveform
seen by node $j$ at any time $t$ is
\begin{eqnarray} \label{eq:timesync-aggwaveform}
A^{c_{1}}_{j,N}(t) = \sum_{i=1}^{N} \frac{ A_{max}K_{j,i}}{N}
p(t-\tau_{o}-T_{i}),
\end{eqnarray}
where $A^{c_{1}}_{j,N}(t)$ is the waveform seen at node $j$
written in the time scale of $c_{1}$ and $A_{i} = A_{max}/N$ for
all $i$. $A_{max}$ is the maximum transmit magnitude of a node.
$T_{i}$ is the random timing offset suffered by the $i$th node,
which encompasses the random clock jitter and estimation error.
This model says that each node $i$'s pulse transmission occurs at
the ideal transmit time $\tau_{0}$ plus some random error $T_{i}$.
In the next section, Section~\ref{sec:pulseProperties}, we find
properties for $T_{i}$ that will give us desirable properties in
$A^{c_{1}}_{j,N}(t)$. Then, in
Section~\ref{sec:timeSynchronization}, we show that our proposed
steady-state synchronization technique and its associated
pulse-connection function will give us the desired properties.

There are two comments about (\ref{eq:timesync-aggwaveform}) that
we want to make. First, note that even though we sum the
transmissions from all $N$ nodes in
(\ref{eq:timesync-aggwaveform}), we do not assume that node $j$
can hear all nodes in the network.  Recall from the pathloss model
that if we have a multi-hop network, then there will be a nonzero
probability that $K_{j,i}=0$.  Thus, node $j$ will not hear from
the nodes whose transmissions have zero magnitude.  Second, it may
be possible that the nodes are told that there are $\bar{N}=vN$
nodes in the network while the actual number of functioning nodes
is $N$.  In which case, each node will transmit with signal
magnitude $A_{i} = A_{max}/(vN)$ and
(\ref{eq:timesync-aggwaveform}) will have a factor of $1/v$. Other
than for this factor, however, the theoretical results that follow
are not affected.

To model the quality of the reception of $A^{c_{1}}_{j,N}(t)$ by
node $j$, we model the reception of a signal by defining a
threshold $\gamma$.  $\gamma$ is the received signal threshold
required for nodes to perfectly resolve the pulse arrival time. If
the maximum received signal magnitude is less than $\gamma$ then
the node does not make any observations and ignores the received
signal waveform. We assume that $\gamma\ll A_{max}$.

In our work we will assume that $p(t)$ takes on the shape
\begin{eqnarray} \label{eq:poft}
p(t) = \left \{ \begin{array}{ll}
q(t) & \textrm{ $-\tau_{nz} < t < 0$}  \\
0 & \textrm{ $t = 0, t \leq -\tau_{nz}, t \geq \tau_{nz}$} \\
-q(-t) & \textrm{ $0 < t < \tau_{nz}$}
\end{array} \right.
\end{eqnarray}
where $\tau_{nz}>0$ is expressed in terms of $c_{1}$. We assume
$q(t)> 0$ for $t\in (-\tau_{nz}, 0)$, $q(t)\ne 0$ only on $t\in
(-\tau_{nz}, 0)$, $\textrm{sup}_{t} |q(t)| = 1$, and $q(t)$ is
uniformly continuous on $(-\tau_{nz}, 0)$. Thus, we see that
$p(t)$ has at most three jump discontinuities (at $t =
0,-\tau_{nz},\tau_{nz}$). $\tau_{nz}$ should be chosen large
compared to $\max_{i}\sigma_{i}^{2}$, i.e.
$\sigma_{i}^{2}<<\tau_{nz}$, where $\sigma_{i}^{2}$ is the value
of $\sigma^{2}$ translated from the time scale of $c_{i}$ to
$c_{1}$. This way, over each synchronization phase, with high
probability a zero-crossing will occur.  For each node, the
duration in terms of $c_{1}$ of a synchronization phase will be
$2\tau_{nz}$.  Note that we assume $\tau_{nz}$ is a value that is
constant in any consistent time scale. This means that even though
nodes have different clocks, identical pulses are transmitted by
all nodes. We define a pulse to be transmitted at time $t$ if the
pulse makes a zero-crossing at time $t$. Similarly, we define the
\emph{pulse receive (arrival) time} for a node as the time when
the observed waveform first makes a zero-crossing. A
\emph{zero-crossing} is defined for signals that have a positive
amplitude and then transition to a negative amplitude. It is the
time that the signal first reaches zero.

For the exchange of synchronization pulses, we assume that nodes
can transmit pulses and receive signals at the same time.  This
simplifying assumption is not required for the ideas presented
here to hold, but simplifies the presentation.  We mention a way
to relax this assumption in Section~\ref{sec:noSimultTxRx}.

\begin{figure}[!h]
\centerline{\psfig{file=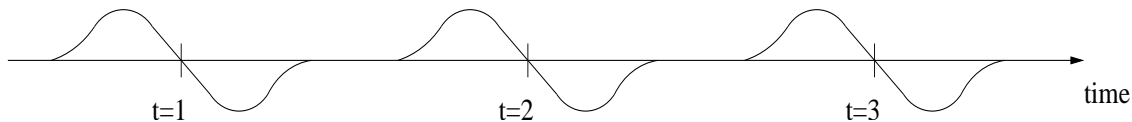,width=15cm,height=1.5cm}}
\vspace{-3mm}
\caption[A pulse train with equispaced zero-crossings.]{\small An
  illustration of a pulse train with equispaced zero-crossings.
  The pulse at each integer value of $t$ is an instance of
  $A_{j,\infty}(t)=\lim_{N\to\infty}A^{c_{1}}_{j,N}(t)$ so we see
  three instances of $A_{j,\infty}(t)$ in the above figure with
  zero-crossings at $t=1,2,3$. We can control the zero-crossings of
  $A_{j,\infty}(t)$ and choose to place it on an integer value of
  $t$.  As a result, we can use these zero-crossings as
  synchronization events since they can be detected simultaneously
  by all nodes in the network.}
\label{fig:pulsetrain}
\end{figure}

In (\ref{eq:timesync-aggwaveform}) and in the discussions above,
we have focused on characterizing the aggregate waveform for any
one synchronization phase. That is,
(\ref{eq:timesync-aggwaveform}) is the waveform seen by any node
$j$ for the synchronization phase centered around node $1$'s
transmission at $t=\tau_{0}$, $\tau_{0}$ a positive integer.  We
can, however, describe a synchronization pulse train in the
following form,
\begin{eqnarray}
\bar{A}^{c_{1}}_{j,N}(t) = \sum_{u=1}^{\infty}\sum_{i=1}^{N}
\frac{A_{max}K_{j,i}}{N} p(t-\tau_{u}-T_{i,u}),
\end{eqnarray}
where $\tau_{u}$ is the integer value of $t$ at the $u$th
synchronization phase, and $T_{i,u}$ is the error suffered by the
$i$th node in the $u$th synchronization phase.  We seek to create
this pulse train with equispaced zero-crossings and use each
zero-crossing as a synchronization event.  An illustration of such
a pulse train is shown in Fig.~\ref{fig:pulsetrain}. For
simplicity, however, most of the theoretical work is carried out
on one synchronization phase.

\subsection{Desired Structural Properties of the Received Signal}\label{sec:pulseProperties}

In this section, we characterize the properties of $T_{i}$ that
give us desirable properties in the aggregate waveform.  From
(\ref{eq:timesync-aggwaveform}), the aggregate waveform seen at
each node $j$ in the network has the form
\begin{eqnarray} \label{eq:waveformGeneralForm}
A_N(t) = \frac 1 N \sum_{i=1}^N A_{max}K_i p(t-\tau_0-T_{i})
\end{eqnarray}
We have dropped the $j$ and $c_{1}$ for notational simplicity
since in this section we deal solely with the received waveform at
a node $j$ in the time scale of $c_{1}$. As we let the number of
nodes grow unbounded ($N\to\infty$), the properties of this limit
waveform can be characterized by Theorem~\ref{theorem:main}. These
properties will be essential for asymptotic cooperative time
synchronization. As a note, in Theorem~\ref{theorem:main} we
present the case for Gaussian distributed $T_{i}$ but similar
results hold for arbitrary zero-mean, symmetrically distributed
$T_{i}$ with finite variance.

\begin{theorem}
\label{theorem:main} Let $p(t)$ be as defined in
equation~(\ref{eq:poft}) and $T_i\sim {\mathcal N}(0,
\frac{\bar{\sigma}^{2}}{\alpha_{i}^{2}})$ with
$\bar{\sigma}^{2}>0$ a constant and
$\frac{\bar{\sigma}^2}{\alpha_{i}^{2}} < B <\infty$ for all $i$,
$B$ a constant. Also, let $K_i$ be defined as in
Section~\ref{sec:prop-model} and be independent from $T_{i}$ for
all $i$. Then, $\lim_{N\to\infty}A_{N}(t) = A_{\infty}(t)$ has the
properties
\begin{itemize}
\item $A_\infty(\tau_0) = 0$, \item $A_\infty(t)>0$ for $t\in
(\tau_{0}-\tau,\tau_{0})$, and $A_\infty(t)<0$ for $t\in
(\tau_{0},\tau_{0}+\tau)$ for some $\tau < \tau_{nz}$. \item
$A_{\infty}(t)$ is odd around $t=\tau_{0}$, i.e.
$A_{\infty}(\tau_{0}+\xi) = -A_{\infty}(\tau_{0}-\xi)$ for
$\xi\geq 0$ \item $A_\infty(t)$ is continuous. \qquad
$\bigtriangleup$
\end{itemize}
\end{theorem}
The properties outlined in Theorem~\ref{theorem:main} will be key
to the synchronization mechanism we describe. The specific value
of $\bar{\sigma}^{2}$ will be determined by our choice of the
pulse-connection function. Before we prove
Theorem~\ref{theorem:main} in Section~\ref{sec:theoremProof} we
develop and motivate a few important related lemmas.

\subsubsection{Polarity and Continuity of $A_\infty(t)$}

At time $t = \tau_1 \ne \tau_0$, we have that
\[ A_N(\tau_1) \;\; = \;\; \sum_{i=1}^N \frac{A_{max}K_i}{N} p(\tau_1-\tau_0-T_{i})
   \;\; = \;\; \sum_{i=1}^N \frac 1 N \bar{M}_i(\tau_1),
\]
where $\bar{M}_i(\tau_1) \stackrel{\Delta}{=} A_{max}K_i
p(\tau_1-\tau_0-T_{i})$. We have the mean of
$\bar{M}_{i}(\tau_{1})$ being
\begin{equation} \label{eqn:MbarMean}
E(\bar{M}_{i}(\tau_{1})) = A_{max}E(K_{i})\int
p(\tau_1-\tau_0-\psi) f_{T_{i}}(\psi) d\psi,
\end{equation}
where $f_{T_{i}}(\psi)$ is the Gaussian pdf
\begin{displaymath}
f_{T_{i}}(\psi) =
\frac{1}{\frac{\bar{\sigma}}{\alpha_{i}}\sqrt{2\pi}}\textrm{exp}
\bigg\{-\frac{(\psi)^2}{2\frac{\bar{\sigma}^{2}}{\alpha_{i}^2}}\bigg\}.
\end{displaymath}
It is clear that the $\bar{M}_{i}(\tau_{1})$'s, for different
$i$'s, do not have the same mean and do not have the same variance
since the two quantities depend on the $\alpha_{i}$ value.  Since
the $\alpha_{i}$'s are characterized by $f_{\alpha}(s)$ (defined
in Section~\ref{sec:systemParameters}), we write the Gaussian
distribution for $T$ as
\begin{displaymath}
f_{T}(\psi,s) =
\frac{1}{\frac{\bar{\sigma}}{s}\sqrt{2\pi}}\textrm{exp}
\bigg\{-\frac{(\psi)^2}{2\frac{\bar{\sigma}^{2}}{s^2}}\bigg\}.
\end{displaymath}
and $\bar{M}_{i}(\tau_{1})$ is in fact a function of $s$ as well,
denoted $\bar{M}_{i}(\tau_{1},s)$. Using $f_{T}(\psi,s)$ and
$\bar{M}_{i}(\tau_{1},s)$, the notation makes it clear that we can
average over the $\alpha_{i}$'s that are characterized by
$f_{\alpha}(s)$. We use the results of
Lemmas~\ref{lemma:polarity_positive} and
\ref{lemma:polarity_negative} to prove the polarity result for
$A_{\infty}(t)$ in Section~\ref{sec:theoremProof}.

\begin{lemma} \label{lemma:polarity_positive}
Given the sequence of independent random variables
$\bar{M}_{i}(\tau_{1})$ with $\tau_{1}<\tau_{0}$,
$E(\bar{M}_{i}(\tau_{1})) = \mu_{i}$, and
$\textrm{Var}(\bar{M}_{i}(\tau_{1})) = \sigma_{i}^{2}$. Then, for
all $i$,
\begin{equation}   \label{eqn:cond3a}
\gamma_{2}>\mu_{i} > \gamma_{1} > 0
\end{equation}
\begin{equation} \label{eqn:cond3b}
\sigma_{i}^{2}<\gamma_{3}<\infty,
\end{equation}
for some constants $\gamma_{1}$, $\gamma_{2}$, and $\gamma_{3}$
and
\begin{displaymath}
\lim_{N\to\infty} \frac{1}{N} \sum_{i=1}^{N} \bar{M}_{i}(\tau_{1}) = \eta(\tau_{1}) > 0
\end{displaymath}
almost surely, where
\begin{eqnarray*}
\eta(\tau_{1}) &=& \int_{\alpha_{low}}^{\alpha_{up}}
E(\bar{M}_{i}(\tau_{1},s)) f_{\alpha}(s) ds  \\ &=&
A_{max}E(K_{i})\int_{\alpha_{low}}^{\alpha_{up}}
\int_{-\infty}^{\infty} p(\tau_1-\tau_0-\psi) f_{T}(\psi,s) d\psi
f_{\alpha}(s) ds. \qquad \bigtriangleup
\end{eqnarray*}
\medskip
\end{lemma}

\begin{lemma} \label{lemma:polarity_negative}
Given the sequence of independent random variables
$\bar{M}_{i}(\tau_{1})$ with $\tau_{1}>\tau_{0}$,
$E(\bar{M}_{i}(\tau_{1})) = \mu_{i}$, and
$\textrm{Var}(\bar{M}_{i}(\tau_{1})) = \sigma_{i}^{2}$.  Then, for
all $i$,
\begin{equation}   \label{eqn:cond4a}
\gamma_{2}<\mu_{i} < \gamma_{1} < 0
\end{equation}
\begin{equation} \label{eqn:cond4b}
\sigma_{i}^{2}<\gamma_{3}<\infty,
\end{equation}
for some constants $\gamma_{1}$, $\gamma_{2}$, and $\gamma_{3}$
and
\begin{displaymath}
\lim_{N\to\infty} \frac{1}{N} \sum_{i=1}^{N} \bar{M}_{i}(\tau_{1}) = \eta(\tau_{1}) < 0
\end{displaymath}
almost surely,  where
\begin{displaymath}
\eta(\tau_{1}) = \int_{\alpha_{low}}^{\alpha_{up}}
E(\bar{M}_{i}(\tau_{1},s)) f_{\alpha}(s) ds. \qquad \bigtriangleup
\end{displaymath}
\medskip
\end{lemma}

The results of Lemma~\ref{lemma:polarity_positive} and
Lemma~\ref{lemma:polarity_negative} are intuitive since given that
$p(t)$ is odd and the Gaussian noise distribution is symmetric, it
makes sense for $A_{\infty}(t)$ to have properties similar to an
odd waveform. Since the proofs of the two lemmas are very similar,
we only prove Lemma~\ref{lemma:polarity_positive}. The proof can
be found in the appendix.

Knowing only the polarity of $A_{\infty}(t)$ is not entirely
satisfying since we would also expect that the limiting waveform
be continuous. The proof of Lemma~\ref{lemma:continuity1} is once
again left for the appendix.

\begin{lemma}
\label{lemma:continuity1}
Using $p(t)$ in (\ref{eq:poft}),
\[ A_\infty(t) \;\; = \;\; \lim_{N\to\infty}\frac{1}{N}\sum_{i=1}^{N} A_{max}K_i
p(t-\tau_0-T_i) \;\; = \;\; \lim_{N\to\infty}\frac{1}{N}\sum_{i=1}^{N}
\bar{M}_{i}(t) \;\; = \;\; \eta(t)
\]
is a continuous function of $t$, where
\begin{eqnarray*}
\eta(t) &=& \int_{\alpha_{low}}^{\alpha_{up}} E(\bar{M}_{i}(t,s))
f_{\alpha}(s) ds \\
&=& A_{max}E(K_{i})\int_{\alpha_{low}}^{\alpha_{up}}
\int_{-\infty}^{\infty} p(t-\tau_0-\psi) f_{T}(\psi,s) d\psi
f_{\alpha}(s) ds.  \qquad \bigtriangleup
\end{eqnarray*}
\medskip
\end{lemma}

\subsubsection{Proof of Theorem~\ref{theorem:main}}
\label{sec:theoremProof}

We can proceed in a
straightforward manner to show that $A_\infty(\tau_0)=0$.  For
$t=\tau_{o}$,
\begin{eqnarray*}
A_N(\tau_0) \;\; = \;\; \sum_{i=1}^N
\frac{A_{max}K_i}{N}p(\tau_0-\tau_0-T_{i}) \;\; = \;\; \frac 1 N
\sum_{i=1}^N A_{max}K_i p(-T_{i})
  \;\; = \;\;  \frac 1 N \sum_{i=1}^N M_i,
\end{eqnarray*}
where $M_i \triangleq -A_{max}K_ip(T_i)$.

Since our goal is to apply some form of the strong law of large
numbers, we first examine the mean of $M_i$.  We have that
$E(M_i)$ $=$ $-A_{max}E(K_i)E(p(T_i))$.  Furthermore,
\[
E(p(T_i)) = \int_{-\infty}^\infty p(\psi)f_{T_i}(\psi)d\psi = 0,
\]
since $p(\psi)$ is odd and $f_{T_{i}}(\psi)$ is
even because it is zero-mean Gaussian.  Thus, $E(M_{i})=0$.

We next consider the variance of $M_i$:
\begin{eqnarray*}
\textrm{Var}(M_i) & = & E(M_i^{2})-E^2(M_i)
  =   A_{max}^2 E(K_i^2p^2(T_i)) \\
& = & A_{max}^2 E(K_i^2)E(p^2(T_i))
  <   A_{max}^2
  <   \infty,
\end{eqnarray*}
where we have used the fact that $E(K_{i}^{2})\leq 1$ and
$|p(t)|\leq 1$.

From the preceding discussion we see that the $M_i$'s are a sequence of
zero mean, finite (but possibly different) variance random variables.
From Stark and Woods~\cite{stark-woods}, we know that if
$\sum_{i=1}^{\infty} \textrm{Var}(M_i)/i^2 < \infty$,
then we have strong convergence of the $M_i$'s:
\[
   \frac 1 N \sum_{i=1}^N M_i \to E(M_i),
\]
with probability-1 as $N\to\infty$.  But it is easy to see that
\[
\sum_{i=1}^\infty \frac{\textrm{Var}(M_i)}{i^2} <
\sum_{i=1}^\infty \frac{A_{max}^2}{i^2} = A_{max}^{2}
\frac{\pi^{2}}{6} < \infty,
\]
so the condition is satisfied.  As a result,
\[
A_N(\tau_0) = \frac 1 N \sum_{i=1}^N M_i \to 0,
\]
as $N \to \infty$.

We have that $A_{\infty}(t)$ is continuous from
Lemma~\ref{lemma:continuity1}.  Thus, next we need to show that
$A_\infty(t)>0$ for $t\in (\tau_{0}-\tau,\tau_{0})$, and
$A_\infty(t)<0$ for $t\in (\tau_{0},\tau_{0}+\tau)$ for some
$\tau<\tau_{nz}$.  We show the case for
$t=\tau_{1}\in(\tau_{0}-\tau,\tau_{0})$ by simply applying
Lemma~\ref{lemma:polarity_positive}.  Since
Lemma~\ref{lemma:polarity_positive} holds for all
$\tau_{1}<\tau_{0}$, there clearly exists a $\tau$ such
$A_{\infty}(t)>0$ for $t\in (\tau_{0}-\tau,\tau_{0})$.  The case
for $t\in(\tau_{0},\tau_{0}+\tau)$ comes similarly from
Lemma~\ref{lemma:polarity_negative}.

Lastly, it remains to be shown that $A_{\infty}(t)$ is odd around
$t=\tau_{0}$.  This, however, is evident from the form of $\eta(t)$.
Since $f_{T}(\psi,s)$ is even in $\psi$ about $0$ and $p(\psi)$
is odd about $0$, it is clear that
$\int_{\infty}^{\infty} p(t-\tau_0-\psi) f_{T}(\psi,s) d\psi$
as a function of $t$ is odd about $\tau_{0}$.  Thus, $\eta(t)$ is odd around
$\tau_{0}$.  This then completes the proof
for Theorem~\ref{theorem:main}. \qquad $\bigtriangleup$

\section{Asymptotic Time Synchronization}
\label{sec:timeSynchronization}

\subsection{The Use of Estimators in Time Synchronization}
\label{assump-theorems}

In this work we want to show that as we let $N\to\infty$ then we
can recover deterministic parameters that allow for time
synchronization.  Such a result would provide rigorous theoretical
support for a new trade-off between network density and
synchronization performance.  To simplify the study, we focus on
the steady-state time synchronization properties of asymptotically
dense networks. In particular, we develop a cooperative technique
that constructs a sequence of equispaced zero-crossings seen by
all nodes which allows the network to maintain time
synchronization indefinitely given that the nodes start with a
collection of equispaced zero-crossings. Starting with a few
equispaced zero-crossings allows us to avoid the complexities of
starting up the synchronization process but still allows us to
show that spatial averaging can be used to average out timing
errors.  If we are able to maintain indefinitely a sequence of
equispaced zero-crossing using cooperative time synchronization,
then it means that spatial averaging can average out all
uncertainties in the system as we let node density grow unbounded.
This recovery of deterministic parameters is our desired result.
Here, we overview the estimators needed for cooperative time
synchronization.

Let $t_{n,i}^{c_{k}}$ be the time, with respect to clock $c_{k}$,
that the $i$th node sees its $n$th pulse. In dealing with the
steady-state properties, we start by assuming that each node $i$
in the network has observed a sequence of $m$ pulse arrival times,
$t_{n-1,i}^{c_{i}},\dots,t_{n-m,i}^{c_{i}}$, that occur at integer
values of $t$, $m$ is an integer. Recall that
$t_{n-1,i}^{c_{i}},\dots,t_{n-m,i}^{c_{i}}$ is defined as a set of
$m$ pulse arrival times in the time scale of $c_{i}$. Therefore,
even though $t_{n-1,i}^{c_{i}},\dots,t_{n-m,i}^{c_{i}}$ occur at
integer values of $t$ (the time scale of $c_{1}$), these values
are not necessarily integers since they are in the time scale of
$c_{i}$. Note also that in our model the pulse arrival time is a
zero-crossing location. Using these $m$ pulse arrival times, each
node $i$ has two distinct, yet closely related tasks. The first
task is time synchronization.  To achieve time synchronization,
node $i$ wants to use these $m$ pulse arrival times to make an
estimate of when the next zero-crossing will occur.  If it can
estimate this next zero-crossing time, then it can effectively
estimate the next integer value of $t$. This estimator can then be
extended to estimate arbitrary times in the future which gives
node $i$ the ability to synchronize to node $1$. The second task
is that node $i$ needs to transmit a pulse so that the sum of all
pulses from the $N$ nodes in the network will create an aggregate
waveform that, in the limit as $N\to\infty$, will give a
zero-crossing at the next integer value of $t$.  This second task
is very significant because if the aggregate waveform gives the
exact location of the next integer value of $t$, then each node
$i$ in the network can use this new zero-crossing along with
$t_{n-1,i}^{c_{i}},\dots,t_{n-m+1,i}^{c_{i}}$ to form a set of $m$
zero-crossing locations.  This new set can then be used to predict
the next zero-crossing location as well as node $i$'s next pulse
transmission time. Recall that determining the pulse transmission
time is the job of the pulse-connection function
$X_{n,i}^{c_{i}}$. With such a setup, synchronization would be
maintained indefinitely.  The zero-crossings that always occur at
integer values of $t$ would provide node $i$ a sequence of
synchronization events and also illustrate how cooperation is
averaging out all random errors.

\begin{figure}[!h]
\centerline{\psfig{file=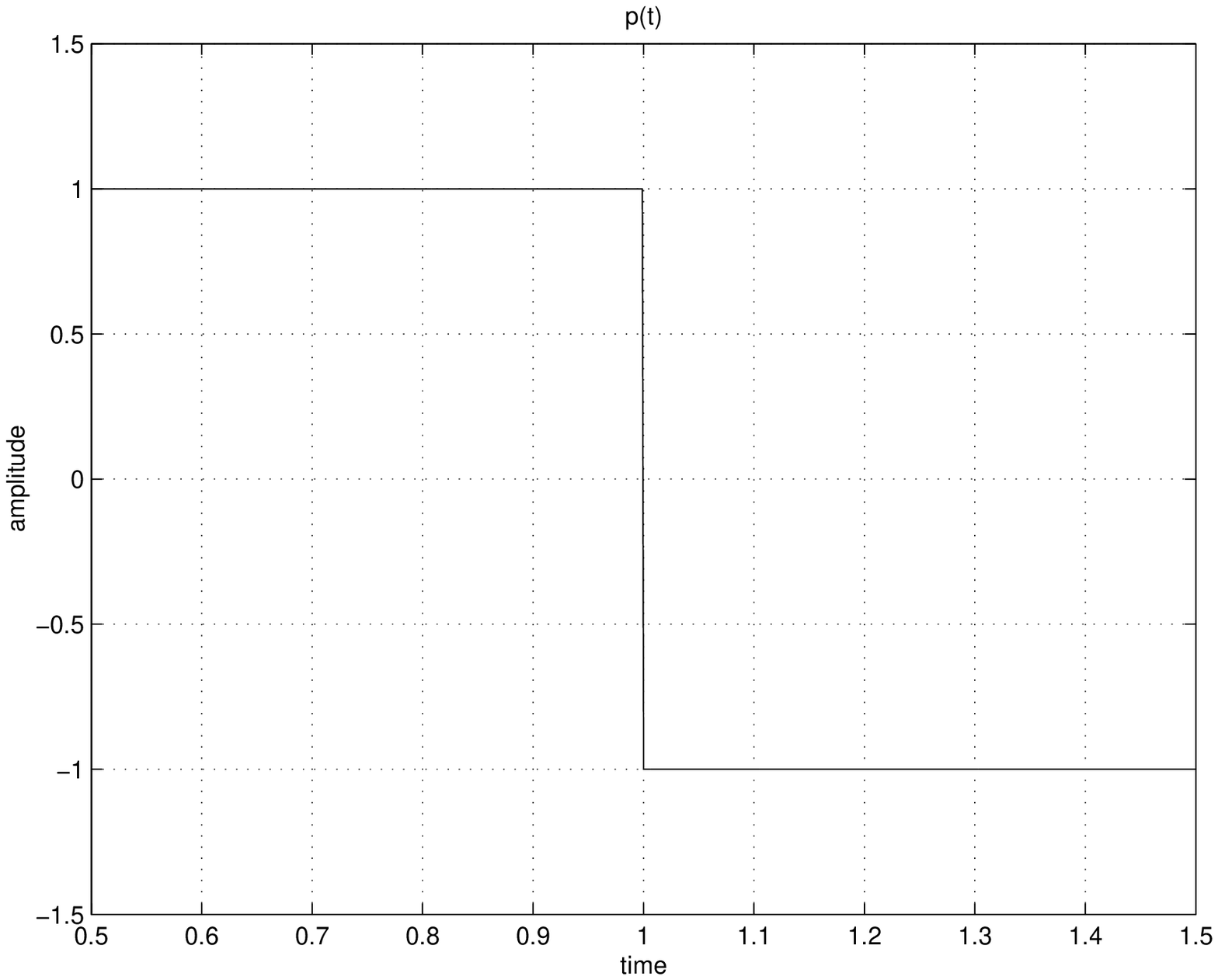,height=5cm}
            \psfig{file=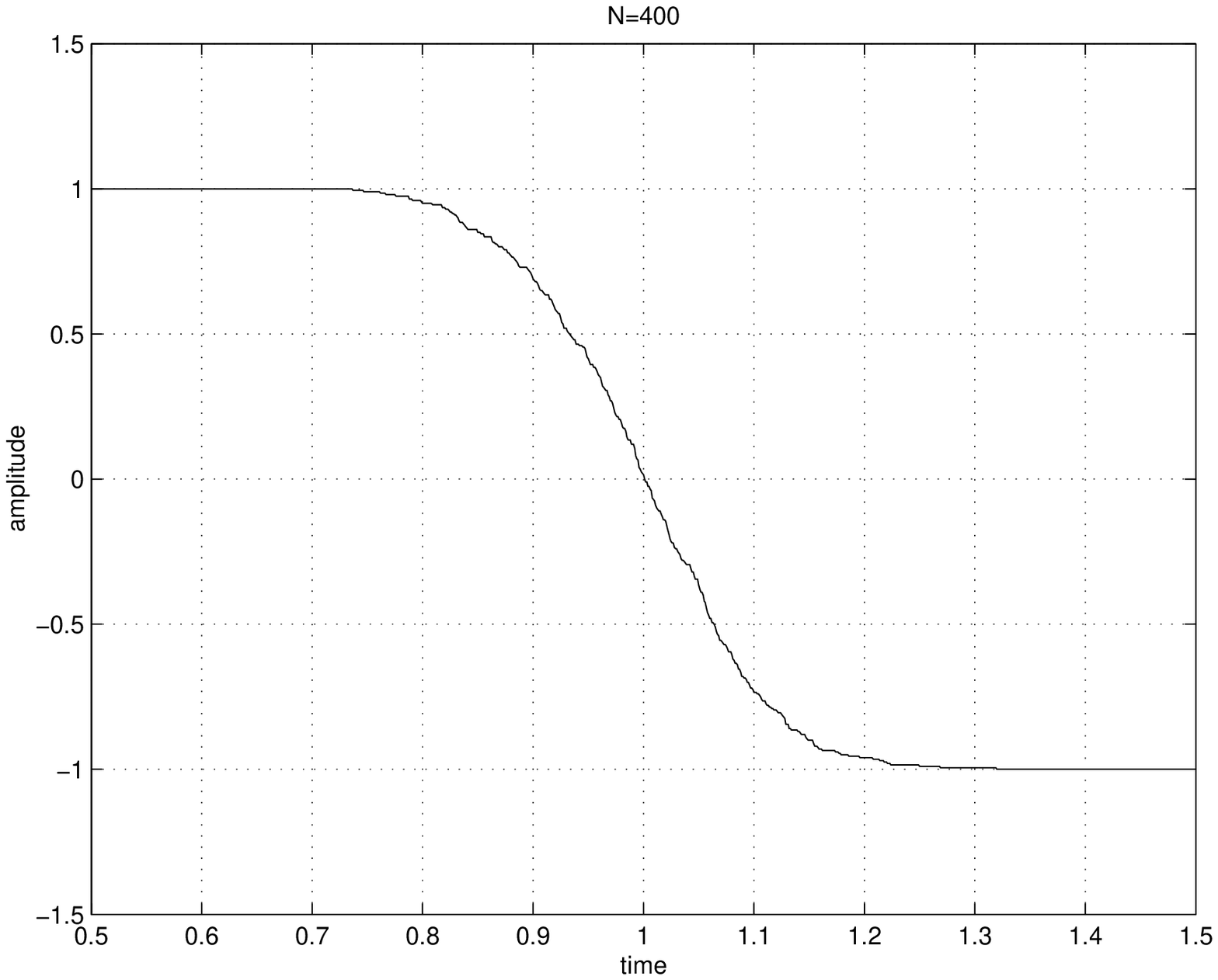,height=5cm}}
\caption[An illustration of the main idea of
Lemma~\ref{theorem1}.]{\small Theorem~\ref{theorem:main} is key in
explaining the intuition first illustrated in
Fig.~\ref{fig:why-sync-holds}.  The pulse $p(t)$ is shown on the
left figure, with
  $\tau_0=1$ and $A_{max}=1$.  On the right we have a realization of $A_N(t)$ ($N=400$),
  and we assume that $K_{j,i}=1$  (no path loss) and $T_{i}\sim {\mathcal N}(0,0.01)$ for all $i$.  As expected from
  Theorem~\ref{theorem:main}, we notice that
  the zero-crossing of the simulated waveform is almost exactly at $t=1$.} \label{fig:example}
\end{figure}

The waveform properties detailed in Theorem~\ref{theorem:main}
play a central role in accomplishing the nodes' task of cooperatively
generating an aggregate waveform with a zero-crossing at the next
integer value of $t$.  From~(\ref{eq:waveformGeneralForm}), if the
arrival time of any pulse at a node $j$ is a random variable of the
form $\tau_0+T_{i}$, where $\tau_{0}$ is the next integer value of
$t$ and $T_{i}$ is zero-mean Gaussian (or in general any symmetric
random variable with zero-mean and finite variance), then
Theorem~\ref{theorem:main} tells us that the aggregate waveform
will make a zero-crossing at the next integer value of $t$.  This
idea is illustrated in Fig.~\ref{fig:example}.

Thus, for achieving time synchronization in an asymptotically
dense network we need to address two issues.  First, we need to
develop an estimator for the next integer value of $t$ given a
sequence of $m$ pulse arrival times that occur at integer values
of $t$. We will call this the {\em time synchronization estimator}
and let us write $V_{n,i}^{c_{i}}$ as the time synchronization
estimator that determines the time, in the time scale of $c_{i}$,
when node $i$ predicts it will see its $n$th zero-crossing. Two,
we need to develop the pulse-connection function $X_{n,i}^{c_{i}}$
such that node $i$'s transmitted pulse will arrive at a node $j$
with the random properties described in
Theorem~\ref{theorem:main}.

\subsection{Time Synchronization Estimator Performance Measure} \label{sec:opt-cond}

Here we establish the conditions for estimating the next pulse
arrival time, or equivalently the next integer value of $t$, given
$m$ pulse arrival times. These conditions apply most directly to
the time synchronization estimator $V_{n,i}^{c_{i}}$ since we want
to synchronize in some desired manner.  The problem of
synchronization is the challenge of having the $i$th node
accurately and precisely predict when the next integer value of
$t$ will occur. In our setup, the reception of a pulse by node $i$
tells it of such an event.

Let us explicitly model the time at an integer value of $t$ in
terms of the clock of node $i$.  Assume $\tau_{0}$ is an integer
value of $t$ and at this time, node $i$ will observe its $n$th
pulse.  Thus, from (\ref{eq:clock}) we have that
\begin{equation} \label{eq:state-eqns}
t_{n,i}^{c_{i}} =
\alpha_{i}(\tau_{0}-\bar{\Delta}_{i})+\Psi_{i}(\tau_{0}).
\end{equation}
The equation makes use of the clock model of node $i$
(\ref{eq:clock}) to tell us the time at clock $c_{i}$ when node
$1$ is at $\tau_{0}$, where $\tau_{0}$ is an integer in the time
scale of $c_{1}$. We are also starting with the assumption that
the zero-crossing that occurs at an integer value of $t$ is
observed by node $i$ at this time.

From (\ref{eq:state-eqns}) we see that the pulse receive time at
node $i$, $t_{n,i}^{c_{i}}$, is a Gaussian random variable whose
mean is parameterized by the unknown vector $\vartheta =
[\alpha_i, \tau_{0}, \bar{\Delta}_{i}]$. Thus, to achieve
synchronization node $i$ will try to estimate the random variable
$t_{n,i}^{c_{i}}$ using a series of $m$ pulse receive times as
observations (recall that $m$ is known).  Note that the
observations are also random variables with distributions
parameterized by $\vartheta$. We want the time synchronization
estimator of node $i$ to make an estimate of $t_{n,i}^{c_{i}}$,
denoted $\hat{t}_{n,i}^{c_{i}}(t_{n-1,i}^{c_{i}},
t_{n-2,i}^{c_{i}},\dots,t_{n-m,i}^{c_{i}})$ which is a function of
past observations $t_{n-1,i}^{c_{i}},
t_{n-2,i}^{c_{i}},\dots,t_{n-m,i}^{c_{i}}$, that meets the
following criteria:
\begin{equation}
E_{\vartheta}\big[\hat{t}_{n,i}^{c_{i}}(t_{n-1,i}^{c_{i}},
t_{n-2,i}^{c_{i}},\dots,t_{n-m,i}^{c_{i}})\big] =
E_{\vartheta}(t_{n,i}^{c_{i}}) \label{eq:opt1}
\end{equation}
\vspace{-5mm}
\begin{equation}
\textrm{argmin}_{\hat{t}_{n,i}^{c_{i}}}
E_{\vartheta}\big[(\hat{t}_{n,i}^{c_{i}}(t_{n-1,i}^{c_{i}},
t_{n-2,i}^{c_{i}},\dots,t_{n-m,i}^{c_{i}})-t_{n,i}^{c_{i}})^{2}\big]
\label{eq:opt2}
\end{equation}
for all $\vartheta$. The subscript $\vartheta$ means that the
expectation is taken over the distributions involved given any
possible $\vartheta$.  The first condition comes from the fact
that given a finite $m$, it is reasonable to want the expected
value of the estimate to be the expected value of the random
variable being estimated for all $\vartheta$. As in the
justification for unbiased estimators, this condition eliminates
unreasonable estimators so that the chosen estimator will perform
well, on average, for all values of $\vartheta$~\cite{Poor:94}.
The second condition is the result of seeking to minimize the mean
squared error between the estimate and the random variable being
estimated for all $\vartheta$.

\subsection{Time Synchronization Estimator} \label{sec:timeSyncEstimator}

For the time synchronization estimator, node $i$ will seek to
estimate $t_{n,i}^{c_{i}}$ given
$t_{n-1,i}^{c_{i}},\dots,t_{n-m,i}^{c_{i}}$. From
(\ref{eq:state-eqns}), we see that ${\mathbf
T}=[t_{n-m,i}^{c_{i}},\dots,t_{n-1,i}^{c_{i}}]^{T}$ is a jointly
Gaussian random vector parameterized by $\vartheta$. Recall that
we  assume $\Psi_{i}(t)$ is a zero mean Gaussian process with
independent and identically distributed samples $\Psi_{i}(t)\sim
{\mathcal N}(0,\sigma^{2})$, for any $t$. Also, since we're
assuming that the zero-crossings at node $i$ occur at consecutive
integer values of $t$, the random variable $t_{n-m,i}^{c_{i}}$ is
Gaussian with $t_{n-m,i}^{c_{i}}\sim {\mathcal
N}(\alpha_{i}(\tau_{0}-m-\bar{\Delta}_{i}), \sigma^{2})$ for some
$\vartheta = [\alpha_i, \tau_{0}-m, \bar{\Delta}_{i}]$.  We also
notice that
\begin{displaymath}
E_{\vartheta}(t_{n-m+1,i}^{c_{i}})=\alpha_{i}(\tau_{0}-m
+1-\bar{\Delta}_{i})=\alpha_{i}(\tau_{0}-m-
\bar{\Delta}_{i})+\alpha_{i}.
\end{displaymath}
Since each noise sample is independent, we see that the
distribution of $\mathbf{T}$ parameterized by $\vartheta$ can be
written as ${\mathbf T} \sim {\mathcal N}({\mathbf M},\Sigma)$
where
\begin{displaymath}
{\mathbf M} = \left[ \begin{array}{c}
\alpha_{i}(\tau_{0}-m-\bar{\Delta}_{i})\\
\alpha_{i}(\tau_{0}-m-\bar{\Delta}_{i})+\alpha_{i}\\
\alpha_{i}(\tau_{0}-m-\bar{\Delta}_{i})+2\alpha_{i}\\
\vdots\\
\alpha_{i}(\tau_{0}-m-\bar{\Delta}_{i})+(m-1)\alpha_{i}
\end{array} \right]
\end{displaymath}
and $\Sigma = \sigma^{2}{\mathbf I}$.

As a result, for any $m$ consecutive observations, we can simplify
notation by using the model
\begin{equation} \label{eq:simple-obs}
{\mathbf Y} = {\mathbf H}{\mathbf \theta} + {\mathbf W},
\end{equation}
where ${\mathbf Y} = [Y_{1} \quad Y_{2} \dots Y_{m}]^{T} =
[t_{n-m,i}^{c_{i}} \quad t_{n-m+1,i}^{c_{i}} \dots
t_{n-1,i}^{c_{i}}]^{T}$ and
\begin{displaymath}
{\mathbf \theta} = \left[ \begin{array}{c}
\theta_{1}\\
\theta_{2}
\end{array} \right] =
\left[ \begin{array}{c}
\alpha_{i}(\tau_{0}-m-\bar{\Delta}_{i})\\
\alpha_{i}
\end{array} \right]
\end{displaymath}
with
\begin{displaymath}
{\mathbf H} = \left[ \begin{array}{ccccc}
1 & 1 & 1 & \ldots & 1\\
0 & 1 & 2 & \ldots & m-1
\end{array} \right]^T
\end{displaymath}
and ${\mathbf W} = [W_{1}\dots W_{m}]^{T}$. Since $\Psi_{i}(t)$ is
a Gaussian noise process, ${\mathbf W} \sim {\mathcal
N}(0,\Sigma)$ with $\Sigma = \sigma^{2}{\mathbf I}$.

Using the simplified notation in (\ref{eq:simple-obs}), we want to
estimate $Y_{m+1}$, where $Y_{m+1}$ is jointly distributed with
${\mathbf Y}$ as
\begin{displaymath}
\left[ \begin{array}{c}
{\mathbf Y}\\
Y_{m+1}
\end{array} \right] \sim {\mathcal N}(
\left[ \begin{array}{c}
{\mathbf M}\\
\theta_{1}+m\theta_{2}
\end{array} \right],
\left[ \begin{array}{cc}
\Sigma & 0\\
0 & \sigma^{2}
\end{array} \right]).
\end{displaymath}
Using this notation, we can rewrite the synchronization criteria
as:
\begin{equation} \label{eq:opt1-simple}
E_{\theta}\big[\hat{Y}_{m+1}(Y_{1}, Y_{2}, \dots,
Y_{m})\big]=E_{\theta}(Y_{m+1})
\end{equation}
\vspace{-5mm}
\begin{equation} \label{eq:opt2-simple}
\textrm{argmin}_{\hat{Y}_{m+1}}
E_{\theta}\big[(\hat{Y}_{m+1}(Y_{1}, Y_{2}, \dots, Y_{m}) -Y_{m+1}
)^{2}\big],
\end{equation}
where $\hat{Y}_{m+1}$ is the estimator for $Y_{m+1}$.

Condition (\ref{eq:opt1-simple}) implies that our estimate must be
unbiased.  Condition (\ref{eq:opt2-simple}) is equivalent to
\begin{displaymath}
\textrm{argmin}_{\hat{Y}_{m+1}}
E_{\theta}\big[(\hat{Y}_{m+1}(Y_{1}, Y_{2}, \dots, Y_{m})
-(\theta_{1}+m\theta_{2}) )^{2}\big].
\end{displaymath}
To see this equivalence, note that
\begin{eqnarray}
\lefteqn{E_{\theta}\big[(\hat{Y}_{m+1}(Y_{1}, Y_{2}, \dots, Y_{m})-Y_{m+1} )^{2}\big]} \nonumber \\
& = & E_{\theta}\big[(\hat{Y}_{m+1}(Y_{1}, Y_{2}, \dots, Y_{m})-(\theta_{1}+m\theta_{2})-W_{m+1} )^{2}\big] \nonumber \\
& = & E_{\theta}\big[(\hat{Y}_{m+1}(Y_{1}, Y_{2}, \dots, Y_{m})-
(\theta_{1}+m\theta_{2}) )^{2}\big]+E\big[ W_{m+1}^{2}\big],
\end{eqnarray}
where the last inequality follows from the independence of
$W_{m+1}$ from all other noise samples. Since the distribution of
of $W_{m+1}$ is independent of $\theta$,
\begin{eqnarray}
\lefteqn{\textrm{argmin}_{\hat{Y}_{m+1}} E_{\theta}\big[(\hat{Y}_{m+1}(Y_{1}, Y_{2}, \dots, Y_{m})-Y_{m+1})^{2}\big]} \nonumber \\
&=& \textrm{argmin}_{\hat{Y}_{m+1}}
E_{\theta}\big[(\hat{Y}_{m+1}(Y_{1}, Y_{2}, \dots, Y_{m})-
(\theta_{1}+m\theta_{2}) )^{2}\big] \nonumber.
\end{eqnarray}
With these two conditions, from~\cite{Poor:94} we see that the
desired estimate for $Y_{m+1}$ will be the uniformly minimum
variance unbiased (UMVU) estimator for
$E_{\theta}(Y_{m+1})=\theta_{1}+m\theta_{2}$.

Using the above linear model, from~\cite{Kay:93} we know the
maximum likelihood (ML) estimate of $\theta$, $\hat{\theta}_{ML}$,
is given by
\begin{equation} \label{eq:thetaHat}
\hat{\theta}_{ML}=({\mathbf H}^{T}\Sigma^{-1} {\mathbf
H})^{-1}{\mathbf H}^{T}\Sigma^{-1}{\mathbf Y}=
(\mathbf{H}^{T}\mathbf{H})^{-1}\mathbf{H}^{T}{\mathbf Y}.
\end{equation}
This estimate achieves the Cramer Rao lower bound, hence is
efficient.  The Fisher information matrix is
$I(\theta)=\frac{{\mathbf H}^{T}{\mathbf H}}{\sigma^{2}}$ and
$\hat{\theta}_{ML} \sim {\mathcal N}(\theta, \sigma^{2}({\mathbf
H}^{T}{\mathbf H})^{-1})$. This means that $\hat{\theta}_{ML}$ is
UMVU.

Again from~\cite{Kay:93}, the invariance of the ML estimate tells
us that the ML estimate for
$\phi=g(\theta)=\theta_{1}+m\theta_{2}$ is
$\hat{\phi}_{ML}=\hat{\theta_{1}}_{ML}+m\hat{\theta_{2}}_{ML}$.
First, it is clear that $\hat{\phi}_{ML} = {\mathbf
C}\hat{\theta}_{ML}$, where ${\mathbf C}=[1\quad  m]$. As a
result, we first see that $E_{\theta}(\hat{\phi}_{ML})={\mathbf
C}E_{\theta}(\hat{\theta}_{ML})=\theta_{1}+m\theta_{2}$ so
$\hat{\phi}_{ML}$ is unbiased. Next, to see that $\hat{\phi}_{ML}$
is also minimum variance we compare its variance to the lower
bound.
\begin{displaymath}
\textrm{Var}_{\theta}(\hat{\phi}_{ML})={\mathbf
C}\sigma^{2}({\mathbf H}^{T}{\mathbf H})^{-1}{\mathbf C}^{T} =
\frac{2\sigma^{2}(2m+1)}{m(m-1)}.
\end{displaymath}
The extension of the Cramer Rao lower bound in~\cite{Kay:93} to a
function of parameters tells us that
\begin{displaymath}
E_{\theta}(\|\hat{g}-g(\theta)\|^{2})\geq {\mathbf
G}(\theta){\mathbf I}^{-1}(\theta){\mathbf G}^{T}(\theta)
\end{displaymath}
with ${\mathbf G}(\theta) = (\nabla_{\theta}g(\theta))^{T}$. In
this case, ${\mathbf G}(\theta)=[1 \quad m]$ so the lower bound to
the mean squared error is
\begin{displaymath}
{\mathbf G}(\theta){\mathbf I}^{-1}(\theta){\mathbf G}^{T}(\theta)
= \frac{2\sigma^{2}(2m+1)}{m(m-1)}.
\end{displaymath}
As a result, we see that $\hat{\phi}_{ML}$ is UMVU.  Since
$\hat{\phi}_{ML}$ is the desired estimate of where the next pulse
arrival time will be, it is the time synchronization estimator.
Thus,
\begin{equation} \label{eq:timesync-estimator}
V_{n,i}^{c_{i}}({\mathbf Y}) = {\mathbf
C}(\mathbf{H}^{T}\mathbf{H})^{-1}\mathbf{H}^{T}{\mathbf Y}.
\end{equation}
Note that
\begin{equation} \label{eq:timesync-estimator-distribution}
V_{n,i}^{c_{i}}({\mathbf Y})=\hat{\phi}_{ML} \sim {\mathcal
N}\Big(\phi,\frac{2\sigma^{2}(2m+1)}{m(m-1)}\Big).
\end{equation}
has a variance that goes to zero as $m\to\infty$.

\subsection{Time Synchronization with No Propagation Delay}
\label{sec:one-hop}

We now need to develop the pulse-connection function so that the
conditions for $T_{i}$ in Theorem~\ref{theorem:main} are
satisfied. Recall we are developing the synchronization technique
under the assumption of no propagation delay, i.e. $\delta(d)=0$.
Given a sequence of $m$ pulse arrival times, the time
synchronization estimator $V_{n,i}^{c_{i}}$ given in
(\ref{eq:timesync-estimator}) gives each node the ability to
predict the next integer value of $t$.  What remains to be
considered is the second part of the synchronization process:
developing a pulse-connection function $X_{n,i}^{c_{i}}$ such that
the aggregate waveform seen by a node $j$ will have the properties
described in Theorem~\ref{theorem:main}.

Let us first consider the distribution of $V_{n,i}^{c_{i}}$.  From
(\ref{eq:timesync-estimator-distribution}), we have that
\begin{displaymath}
V_{n,i}^{c_{i}}({\mathbf Y}) \sim {\mathcal N}
\bigg(\alpha_{i}(\tau_{0}-m-\bar{\Delta}_{i})+m\alpha_{i},\frac{2\sigma^{2}(2m+1)}{m(m-1)}\bigg).
\end{displaymath}
Using (\ref{eq:clock}), we can translate $V_{n,i}^{c_{i}}({\mathbf
Y})$ into the time scale of $c_{1}$ as
\begin{displaymath}
V_{n,i}^{c_{i}}({\mathbf Y}) = \alpha_{i}(V_{n,i}^{c_{1}}({\mathbf
Y})-\bar{\Delta}_{i}) + \Psi_{i}
\end{displaymath}
which gives
\begin{displaymath}
V_{n,i}^{c_{1}}({\mathbf Y}) = \frac{(V_{n,i}^{c_{i}}({\mathbf Y})
- \Psi_{i})}{\alpha_{i}} + \bar{\Delta}_{i}.
\end{displaymath}
This means that
\begin{equation} \label{eq:compToNoSimult}
V_{n,i}^{c_{1}}({\mathbf Y}) \sim {\mathcal
N}\bigg(\tau_{0},\frac{\sigma^{2}}{\alpha_{i}^{2}}\bigg(1+\frac{2(2m+1)}{m(m-1)}\bigg)\bigg).
\end{equation}
Under our assumption of $\delta(d)=0$, any transmission by node
$i$ will be instantaneously seen by any node $j$.  As a result,
the random variable $V_{n,i}^{c_{1}}({\mathbf Y})$ will be seen as
the pulse arrival time at node $j$, in the time scale of $c_{1}$.

Due to the assumption of no propagation delay, defining
$X_{n,i}^{c_{1}}({\mathbf
Y})\stackrel{\Delta}{=}V_{n,i}^{c_{1}}({\mathbf Y})$ will give us
the desired properties in the aggregate waveform.  To see this,
let us compare the distribution of $X_{n,i}^{c_{1}}({\mathbf Y})$
to the assumptions of Theorem~\ref{theorem:main}. Since $\tau_{0}$
is the ideal crossing time in the time scale of $c_{1}$, we have
\begin{displaymath}
X_{n,i}^{c_{1}}({\mathbf Y}) = \tau_{0} + T_{i}.
\end{displaymath}
Therefore, we see that
\begin{equation} \label{eq:errorVariance}
\textrm{Var}(T_{i}) =
\frac{\sigma^{2}}{\alpha_{i}^{2}}\bigg(1+\frac{2(2m+1)}{m(m-1)}\bigg)
= \frac{\bar{\sigma}^{2}}{\alpha_{i}^{2}},
\end{equation}
where $\bar{\sigma}^{2}$ from Theorem~\ref{theorem:main} is
\begin{displaymath} \bar{\sigma}^{2} =
\sigma^{2}\bigg(1+\frac{2(2m+1)}{m(m-1)}\bigg).
\end{displaymath}
We have shown that using the pulse connection function
$X_{n,i}^{c_{1}}({\mathbf
Y})\stackrel{\Delta}{=}V_{n,i}^{c_{1}}({\mathbf Y})$ satisfies the
conditions of Theorem~\ref{theorem:main}.  Thus, all the results
of the theorem apply.

As a result, we have established a time synchronization estimator
$V_{n,i}^{c_{1}}({\mathbf Y})$ and a pulse-connection function
$X_{n,i}^{c_{1}}({\mathbf Y})$.  In the case of $\delta(d)=0$, we
have that $X_{n,i}^{c_{1}}({\mathbf
Y})\stackrel{\Delta}{=}V_{n,i}^{c_{1}}({\mathbf Y})$, or in the
time scale of $c_{i}$, $X_{n,i}^{c_{i}}({\mathbf
Y})\stackrel{\Delta}{=}V_{n,i}^{c_{i}}({\mathbf Y})$.  When each
node in the network uses the pulse-connection function
$X_{n,i}^{c_{i}}({\mathbf Y})$ we have a resulting aggregate
waveform that has a zero-crossing at the next integer value of $t$
as $N\to\infty$. This fact follows from applying
Theorem~\ref{theorem:main}. Thus, we have an asymptotic
steady-state time synchronization method that can maintain a
sequence of equispaced zero-crossings occurring at integer values
of $t$. An interesting feature of this synchronization technique
is that no node needs to know any information about its location
or its surrounding neighbors.

\subsubsection{Cooperation without Simultaneous Transmission and
Reception} \label{sec:noSimultTxRx}

Before ending this section, let us comment on the assumption of
simultaneous transmission and reception. One way to relax this
assumption is to divide the network into two disjoint sets of
nodes, say the odd numbered nodes and the even numbered nodes,
where each set is still uniformly distributed over the area. Then,
the odd nodes and the even nodes will take turns transmitting and
receiving.  For example, the odd numbered nodes can transmit
pulses at odd values of $t$ and the even numbered nodes will
listen. The even numbered nodes will then transmit pulses at the
even values of $t$ and the odd numbered nodes will listen.  With
such a scheme, nodes do not transmit and receive pulses
simultaneously, but can still take advantage of spatial averaging.
The odd numbered nodes will see an aggregate waveform generated by
a subset of the even numbered nodes and the even numbered nodes
will receive a waveform cooperatively generated by the odd
numbered nodes.  Let us take a more detailed look at this scheme.

\begin{figure}[!h]
\centerline{\psfig{file=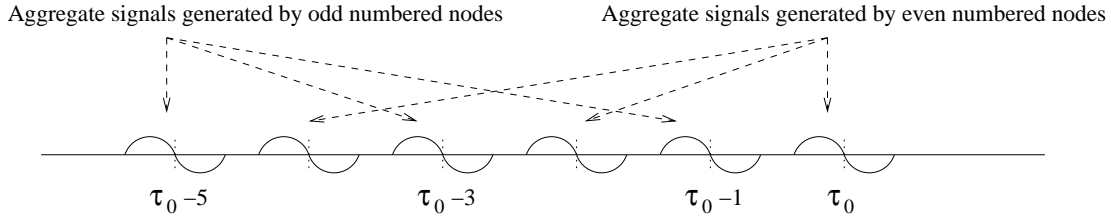,width=15cm}}
\caption{\small In the above figure, we assume $\tau_{0}$ is an
even integer value of $t$ and $m=3$.  Therefore, each even
numbered node will turn on its receiver to receive the aggregate
signal arriving at times $\tau_{0}-5$, $\tau_{0}-3$, and
$\tau_{0}-1$. Using these three received times, it can then
estimate the time of $\tau_{0}$. Thus, the aggregate signal
occurring at $\tau_{0}$ is cooperatively generated by the even
numbered nodes and is received by the odd numbered nodes.}
\label{fig:noSimultTxRx}
\end{figure}

In Fig.~\ref{fig:noSimultTxRx} we assume that $\tau_{0}$ is an
even integer value of $t$ and use $m=3$.  Each even numbered node
will use the aggregate signals occurring at $\tau_{0}-5$,
$\tau_{0}-3$, and $\tau_{0}-1$ to estimate $\tau_{0}$ and
cooperatively the even nodes will generate the aggregate signal at
$\tau_{0}$. The odd numbered nodes will then use the aggregate
signals occurring at $\tau_{0}-4$, $\tau_{0}-2$, and $\tau_{0}$ to
generate the aggregate signal at $\tau_{0}+1$. Therefore, the odd
and even numbered nodes can take turns transmitting and receiving
signals and nodes never need to simultaneously transmit and
receive.

Of course, such a setup would require a modification of the
estimators used by the nodes. Nodes will receive a vector of $m$
observations $\mathbf{Y}$ with $\mathbf{Y}[l+1] =
\alpha_{i}(\tau_{0}+1-2(m-l)-\bar{\Delta}_{i})+\Psi_{i}$ for $l =
0,1,\ldots,m-1$. With such a mechanism, the ${\mathbf H}$ matrix
in equation (\ref{eq:simple-obs}) would change to
\begin{displaymath}
{\mathbf H} = \left[ \begin{array}{ccccc}
1 & 1 & 1 & \ldots & 1\\
0 & 2 & 4 & \ldots & 2(m-1)
\end{array} \right]^T
\end{displaymath} and $\theta$ becomes
\begin{displaymath}
{\mathbf \theta} = \left[ \begin{array}{c}
\theta_{1}\\
\theta_{2}
\end{array} \right] =
\left[ \begin{array}{c}
\alpha_{i}(\tau_{0}+1-2m-\bar{\Delta}_{i})\\
\alpha_{i}
\end{array} \right].
\end{displaymath}

To estimate the location $\tau_{0}$ in the time scale of $c_{i}$,
we can proceed as in Section~\ref{sec:timeSyncEstimator}:
\begin{displaymath}
\hat{\theta}_{ML}=({\mathbf H}^{T}\Sigma^{-1} {\mathbf
H})^{-1}{\mathbf H}^{T}\Sigma^{-1}{\mathbf Y}=
(\mathbf{H}^{T}\mathbf{H})^{-1}\mathbf{H}^{T}Y
\end{displaymath}
will be distributed $\hat{\theta}_{ML} \sim {\mathcal N}(\theta,
\sigma^{2}({\mathbf H}^{T}{\mathbf H})^{-1})$ and
$\hat{\theta}_{ML}$ is UMVU. This leads to the UMVU estimate
$\hat{\phi}_{ML} = {\mathbf C}\hat{\theta}_{ML}$, where ${\mathbf
C}=[1\quad  2m-1]$, and
$E(\hat{\phi}_{ML})={\mathbf
C}E(\hat{\theta}_{ML})=\theta_{1}+(2m-1)\theta_{2}$.
In this case, the variance of $\hat{\phi}_{ML}$ will be
$\textrm{Var}_{\theta}(\hat{\phi}_{ML})={\mathbf
C}\sigma^{2}({\mathbf H}^{T}{\mathbf H})^{-1}{\mathbf C}^{T}$, and
thus we have that
\begin{displaymath}
V_{n,i}^{c_{i}}(\mathbf{Y}) = \hat{\phi}_{ML} \sim
\mathcal{N}\bigg(\alpha_{i}(\tau_{0}+1-2m-\bar{\Delta}_{i})+(2m-1)\alpha_{i},
\frac{\sigma^{2}(2m+1)(2m-1)}{m(m-1)(m+1)}\bigg).
\end{displaymath}
Converted to the time scale of $c_{1}$ we have
\begin{equation} \label{eq:noSimult}
V_{n,i}^{c_{1}}(\mathbf{Y}) \sim \mathcal{N}\bigg(\tau_{0},
\frac{\sigma^{2}}{\alpha_{i}^{2}}\bigg(1+\frac{(2m+1)(2m-1)}{m(m-1)(m+1)}\bigg)\bigg).
\end{equation}
Comparing equations (\ref{eq:compToNoSimult}) and
(\ref{eq:noSimult}), we see that they have the same form.  As a
result, we can again set $X_{n,i}^{c_{i}}({\mathbf
Y})\stackrel{\Delta}{=}V_{n,i}^{c_{i}}({\mathbf Y})$ and achieve
cooperative time synchronization.

\section{Time Synchronization with Propagation Delay}
\label{sec:timeDelay}

We now extend the ideas of cooperative time synchronization to the
situation where signals suffer not only from pathloss but also
propagation delay.  It turns out that the effect of propagation
delay can also be addressed using the concept we have been using
throughout this paper --- averaging out errors using the large
number of nodes in the network.

In this section, we use the pathloss and propagation delay model
detailed in Section~\ref{sec:delay-model}.  We introduce a time
delay function $\delta(d)$.  For generality, we explicitly model a
multi-hop network where we have a $K(d)$ function that is zero for
$d$ greater than some distance $R$, i.e. $K(d) = 0$ for $d>R$.
Such a model implies that the aggregate signal seen at any node
$j$ is influenced only by the set of nodes inside a circle of
radius $R$ centered at node $j$. With this we can effectively
divide the network into two disjoint sets, a set of {\em interior
nodes} and a set of {\em boundary nodes}.  An interior node $j$ is
defined to be a node whose distance from the nearest network
boundary is greater than or equal to $R$.  A boundary node is thus
defined to be a node that is a distance less than $R$ away from
the nearest network boundary.

We make this distinction since the synchronization technique for
each set of nodes is different.  Please note that if a pathloss
function where $K(d) = 0$ for $d>R$ is unreasonable, then we
simply choose $R$ to be infinite and consider all nodes in the
network to be boundary nodes.

Using the propagation delay model, $D_{j,i}$ will obviously modify
the general received aggregate waveform seen at any node $j$.  In
fact, equation (\ref{eq:timesync-aggwaveform}) will now be written
as
\begin{eqnarray} \label{eq:timesync-aggwaveform-withDelay}
A^{c_{1}}_{j,N}(t) & = & \sum_{i=1}^{N} \frac{A_{max}K_{j,i}}{N}
p(t-\tau_{o}-T_{i}-D_{j,i}).
\end{eqnarray}
For $N$ large, this model will give an accurate characterization
of the aggregate waveform seen at node $j$.

\subsection{Conceptual Motivation}
\label{sec:propDelay_motivation}

From equation (\ref{eq:timesync-aggwaveform-withDelay}), it is
clear that the aggregate waveform will not have a zero-crossing at
$\tau_{0}$ for every node $j$ because of the presence of the
$D_{j,i}$ random variables.  Therefore, to average out propagation
delay, the idea we employ is to have each node introduce a
\emph{random} artificial time shift that counteracts the effect of
the time delay random variable. More precisely, we want to
introduce another random variable $D_{fix}$ such that
$D_{fix}+D_{j}$ will have zero mean and a symmetric distribution.
At the same time, we assume each node knows $K(\cdot)$ and
$\delta(\cdot)$ and will also introduce an artificial scaling
factor $K_{fix}=K(\delta^{-1}(-D_{fix}))$ to simplify the analysis
of the aggregate waveform. This means that instead of using the
scaling factor $A_{i}=A_{max}/N$, each node $i$ will scale its
transmitted pulse by $A_{i}=A_{max}K_{fix}/N$. For the motivation
in this section, let us assume that node $j$ is an interior node.

To find the distribution of $D_{fix}$, we consider the following.
$D_{j}$ has density $f_{D_{j}}(x)$ and let $f_{D_{fix}}(x)$ be the
density of $D_{fix}$.  Since $D_{j}$ and $D_{fix}$ are
independent, we know that the density of $D_{T}=D_{fix}+D_{j,i}$,
$f_{D_{T}}(x)$, will be the convolution of $f_{D_{j}}(x)$ and
$f_{D_{fix}}(x)$.  Therefore, by the properties of the convolution
function, if we set
$f_{D_{fix}}(x)\stackrel{\Delta}{=}f_{D_{j}}(-x)$, then we have
that $f_{D_{T}}(x)$ is symmetric, i.e.
$f_{D_{T}}(x)=f_{D_{T}}(-x)$. As well, since $D_{j}$ has finite
expectation, it is easy to see that $E(D_{T}) = 0$.

Given a sequence of $m$ zero-crossings that we know to be
occurring at integers of $t$, we can still use
$V_{n,i}^{c_{1}}({\mathbf Y})$ (from (\ref{eq:timesync-estimator})
in the time scale of node $1$) as the time synchronization
estimator. However, with propagation delay, the pulse-connection
function will now be $X_{n,i}^{c_{1}}({\mathbf
Y})=V_{n,i}^{c_{1}}({\mathbf Y})+D_{fix} =
\tau_{o}+T_{i}+D_{fix}$. With $D_{fix}$ and $K_{fix}$ included, we
can rewrite equation (\ref{eq:timesync-aggwaveform-withDelay}) as
\begin{eqnarray} \label{eq:timesync-aggwaveform-withDelayAndFix}
A^{c_{1}}_{j,N}(t) = \sum_{i=1}^{N}
\frac{A_{max}K_{fix}K_{j,i}}{N}
p(t-\tau_{o}-T_{i}-D_{fix}-D_{j,i}).
\end{eqnarray}
It is important to see that since $D_{j}$ has the same
distribution for \emph{all} interior nodes $j$, equation
(\ref{eq:timesync-aggwaveform-withDelayAndFix}) holds for every
node $j$ that is an interior node.  This means that for the
network to cooperatively generate the waveform in
(\ref{eq:timesync-aggwaveform-withDelayAndFix}) each transmit node
$i$ needs to have the following additional knowledge: (1) the
distribution of $D_{fix}$ whose density is
$f_{D_{fix}}(x)\stackrel{\Delta}{=}f_{D_{j}}(-x)$, where $j$ is an
interior node, and (2) the functions $K(\cdot)$ and
$\delta(\cdot)$ to generate $K_{fix}$. With this knowledge, we can
use equation (\ref{eq:timesync-aggwaveform-withDelayAndFix}) to
study the aggregate waveform seen at any interior node $j$. In
fact, we find that the aggregate waveform has limiting properties
that are similar to those outlined in Theorem~\ref{theorem:main}.
These properties are described in
Theorem~\ref{theorem:main-delay}.

\begin{theorem}
\label{theorem:main-delay} Let $p(t)$ be as defined in
equation~(\ref{eq:poft}) and $T_i\sim {\mathcal N}(0,
\frac{\bar{\sigma}^{2}}{\alpha_{i}^{2}})$ with
$\bar{\sigma}^{2}>0$ a constant and
$\frac{\bar{\sigma}^2}{\alpha_{i}^{2}} < B <\infty$ for all $i$,
$B$ a constant. $K_{j,i}$ and $D_{j,i}$ are defined as in
Section~\ref{sec:delay-model} and $D_{fix}$ with density
$f_{D_{fix}}(x)\stackrel{\Delta}{=}f_{D_{j}}(-x)$ is independent
from $D_{j,i}$.  $K_{fix}=K(\delta^{-1}(-D_{fix}))$ and let
$D_{j,i}$, $D_{fix}$, and $T_{i}$ be mutually independent for all
$i$. Then, for any interior node $j$ with $A_{j,N}^{c_{1}}(t)$ as
defined in (\ref{eq:timesync-aggwaveform-withDelayAndFix}),
$\lim_{N\to\infty}A_{j,N}^{c_{1}}(t) = A_{j,\infty}^{c_{1}}(t)$
has the properties
\begin{itemize}
\item $A_{j,\infty}^{c_{1}}(\tau_0) = 0$, \item
$A_{j,\infty}^{c_{1}}(t)$ is odd around $t=\tau_{0}$, i.e.
$A_{j,\infty}^{c_{1}}(\tau_{0}+\xi) =
-A_{j,\infty}^{c_{1}}(\tau_{0}-\xi)$ for $\xi\geq 0$. \qquad
$\bigtriangleup$
\end{itemize}
\end{theorem}
The proof of Theorem~\ref{theorem:main-delay} is left for the
appendix.

From the arguments so far, it seems that time synchronization with
delay, at least for interior nodes, can be solved simply by
modifying the pulse-connection function $X_{n,i}^{c_{1}}({\mathbf
Y})$ and changing the scaling factor to $A_{i}=A_{max}K_{fix}/N$.
Theorem~\ref{theorem:main-delay} tells us that the limiting
aggregate waveform makes a zero-crossing at the next integer value
of $t$ and the waveform is odd. Thus, we can use this
zero-crossing as a synchronization event and maintain
synchronization in a manner identical to the technique used in the
situation without propagation delay. This, however, unfortunately
is not the case. In order to implement the above concept, we need
to find the random variable, $D_{fix}^{c_{i}}$, in the time scale
of $c_{i}$, that corresponds to $D_{fix}$ such that
\begin{eqnarray*}
(V_{n,i}^{c_{i}}({\mathbf Y})+D_{fix}^{c_{i}})^{c_{1}} & = &
\frac{V_{n,i}^{c_{i}}({\mathbf
Y})+D_{fix}^{c_{i}}-\Psi_{i}}{\alpha_{i}}+\bar{\Delta}_{i} \\
& = & V_{n,i}^{c_{1}}({\mathbf Y}) +
\frac{D_{fix}^{c_{i}}}{\alpha_{i}} \\
& = & V_{n,i}^{c_{1}}({\mathbf Y})+D_{fix}.
\end{eqnarray*}
This means that we need $D_{fix}^{c_{i}}/ \alpha_{i} = D_{fix}$.
However, each node $i$ cannot find $D_{fix}^{c_{i}}$ that
satisfies this since it does not know its $\alpha_{i}$.

\subsection{Time Synchronization of Interior Nodes}
\label{sec:timeSync-Interior}

Since the $i$th node does not know its own value of $\alpha_{i}$,
to do time synchronization with propagation delay we can have each
node estimate its $\alpha_{i}$ value.  However, this estimate will
not be perfect and we may no longer have the symmetric limiting
aggregate waveform described by Theorem~\ref{theorem:main-delay}.
This means that the center zero-crossing might occur some
$\epsilon$ away from $\tau_{0}$, $\tau_{0}$ an integer value of
$t$. However, steady-state time synchronization can be maintained
if the network can use a sequence of $m$ equispaced zero-crossings
that occur at $t=\tau_{0}-m+\epsilon, \tau_{0}-m+1+\epsilon,
\tau_{0}-m+2+\epsilon,\ldots,\tau_{0}-1+\epsilon$, where
$\tau_{0}$ is an integer value of $t$, to cooperatively generate a
limiting aggregate waveform that has a zero-crossing at
$\tau_{0}+\epsilon$. In such a situation, the network will be able
to construct a sequence of equispaced zero-crossings and maintain
the occurrence of these zero-crossings indefinitely. The idea is
the same as in the case without propagation delay, but the only
difference here would be that the zero-crossings do not occur at
integer values of $t$.  Let us give a more formal description of
this idea.

Using notation from Section~\ref{sec:timeSyncEstimator}, we start
with the assumption that each interior node $i$ has a sequence of
$m$ observations that has the form
\begin{equation} \label{eq:observation_form}
\alpha_{i}(\tau_{0}-m+l+\epsilon-\bar{\Delta}_{i})+\Psi_{i},
\end{equation}
where $l=0,1,\ldots,m-1$ and $\epsilon$ is known. To develop the
time synchronization estimator $V_{n,i}^{c_{i}}({\mathbf Y})$ and
the pulse-connection function $X_{n,i}^{c_{i}}({\mathbf Y})$, we
consider the observations made by each node. If we assume that
each node knows the value of $\epsilon$, the vector of
observations can be written as in (\ref{eq:simple-obs})
\begin{displaymath}
{\mathbf Y} = \bar{{\mathbf H}}{\mathbf \theta} + {\mathbf W},
\end{displaymath}
where the matrix $\bar{{\mathbf H}}$ in this case is
\begin{displaymath}
\bar{{\mathbf H}} = \left[ \begin{array}{ccccc}
1 & 1 & 1 & \ldots & 1\\
\epsilon & 1+\epsilon & 2+\epsilon & \ldots & m-1+\epsilon
\end{array} \right]^T.
\end{displaymath}
Using this model, we can follow the development in
Section~\ref{sec:timeSyncEstimator} to find the the time
synchronization estimator
\begin{equation} \label{eq:timesync-estimator-propDelay}
V_{n,i}^{c_{i}}({\mathbf Y},\epsilon) = {\mathbf
C}(\bar{\mathbf{H}}^{T}\bar{\mathbf{H}})^{-1}\bar{\mathbf{H}}^{T}{\mathbf
Y},
\end{equation}
where ${\mathbf C} = [1\quad m]$.  This estimator will give each
node the ability to optimally estimate the next integer value of
$t$. Note that the variance of the time synchronization estimator
is
\begin{equation}
\textrm{Var}_{\theta}(V_{n,i}^{c_{i}}({\mathbf Y},\epsilon)) =
{\mathbf C}\sigma^{2}(\bar{{\mathbf H}}^{T}\bar{{\mathbf
H}})^{-1}{\mathbf C}^{T} =
\sigma^{2}\bigg(\frac{2(2m+1)}{m(m-1)}+\frac{12\epsilon(\epsilon-1-m)}{(m-1)m(m+1)}\bigg).
\end{equation}
Using the time synchronization estimator, we can choose the
pulse-connection function as
\begin{equation} \label{eq:pulse-connection-propDelay}
X_{n,i}^{c_{i}}({\mathbf Y})=V_{n,i}^{c_{i}}({\mathbf
Y},\epsilon)+\hat{\alpha}_{i}D_{fix}=V_{n,i}^{c_{i}}({\mathbf
Y},\epsilon)+D^{c_{i}}_{fix},
\end{equation}
where each time node $i$ makes the estimate
$V_{n,i}^{c_{i}}({\mathbf Y},\epsilon)$ it also estimates
$\hat{\alpha}_{i}$ as
\begin{displaymath}
\hat{\alpha_{i}} = \bar{{\mathbf
C}}(\bar{\mathbf{H}}^{T}\bar{\mathbf{H}})^{-1}\bar{\mathbf{H}}^{T}{\mathbf
Y},
\end{displaymath}
$\bar{{\mathbf C}}=[0\quad 1]$.  We find that $\hat{\alpha}_{i}
\sim {\mathcal N}(\alpha_{i}, 12\sigma^{2}/((m-1)m(m+1)))$. Since,
from Section~\ref{sec:propDelay_motivation}, we know we want
$D_{fix}^{c_{i}}/ \alpha_{i} = D_{fix}$, we have set
$D_{fix}^{c_{i}} \stackrel{\Delta}{=} \hat{\alpha}_{i}D_{fix}$.
Notice that since $D_{fix}^{c_{i}}$ is simply a realization of
$D_{fix}$ multiplied by node $i$'s estimate of $\alpha_{i}$, node
$i$ can use the realization of $D_{fix}$ and find
$K_{fix}=K(\delta^{-1}(-D_{fix}))$.

With our choice of $X_{n,i}^{c_{i}}({\mathbf Y})$ in
(\ref{eq:pulse-connection-propDelay}), we see that
\begin{displaymath}
(V_{n,i}^{c_{i}}({\mathbf Y},\epsilon)+D_{fix}^{c_{i}})^{c_{1}}
\;\;=\;\; V_{n,i}^{c_{1}}({\mathbf Y},\epsilon)+Z_{i}D_{fix}
\;\;=\;\; \tau_{0} +T_{i}+Z_{i}D_{fix},
\end{displaymath}
where $Z_{i}\sim{\mathcal
N}(1,12\sigma^{2}/(\alpha_{i}^{2}(m-1)m(m+1)))$, and
$\tau_{0}+T_{i}=V_{n,i}^{c_{1}}({\mathbf Y},\epsilon)$. Because of
the random factor $Z_{i}$, we see that
$D_{T}=Z_{i}D_{fix}+D_{j,i}$ is no longer a symmetric
distribution.  As a result, the limiting aggregate waveform
\begin{equation} \label{eq:aggregate-waveform-timeDelay}
A_{j,\infty}^{c_{1}}(t)=\lim_{N\to\infty}A^{c_{1}}_{j,N}(t) =
\lim_{N\to\infty}\sum_{i=1}^{N} \frac{A_{max}K_{fix}K_{j,i}}{N}
p(t-\tau_{o}-T_{i}-Z_{i}D_{fix}-D_{j,i})
\end{equation}
may not have a zero-crossing at $t=\tau_{0}$.

Thus, if we can find an $\epsilon$ such that each node $i$ using a
set of observations of the form (\ref{eq:observation_form}) allows
the network to cooperatively generate the waveform in
(\ref{eq:aggregate-waveform-timeDelay}) that has its zero-crossing
occurring at $t=\tau_{0}+\epsilon$ (in the time scale of $c_{1}$),
then we have steady-state time synchronization. This is because
the network would be able to use a sequence of $m$ observations to
generate the next observation that gives the same information as
any of the previous observations.  Thus, by always taking the $m$
most recent observations, the process can continue forever and
maintain synchronization.  Each node $i$ would need to know
distribution of $D_{fix}$, the value of $\epsilon$, and the
functions $K(\cdot)$ and $\delta(\cdot)$.  Therefore, we find that
steady-state time synchronization of the interior nodes is
possible under certain conditions. As a note, no interior node
needs to know any location information.

\subsection{Time Synchronization of Boundary Nodes}
\label{sec:timeSync-Boundary}

Before we consider the synchronization of boundary nodes, we note
that the key requirement for each boundary node $i$ is to have a
pulse-connection function given in equation
(\ref{eq:pulse-connection-propDelay}).  The reason that this must
be the pulse-connection for every boundary node $i$ is because the
analysis for the interior nodes assumes that the aggregate
waveform seen by any interior node $j$ is created by pulse
transmissions occurring at a time determined by
(\ref{eq:pulse-connection-propDelay}).  Since the aggregate
waveform seen by some interior nodes are created by pulse
transmissions from boundary nodes, each boundary node must have
the appropriate pulse-connection function.  This requirement,
however, proves to be extremely problematic and reveals a
limitation of the elegant technique of averaging out timing delay
when we come to boundaries of the network.

The problem comes because $D_{fix}+D_{j,i}$ already does not have
a symmetric distribution if $j$ is a boundary node. Recall that
$f_{D_{fix}}(x) = f_{D_{j}}(-x)$ when $j$ is an interior node and
$f_{D_{j}}(x) = f_{D_{l}}(x)$ when $j$ and $l$ are both interior
nodes. However, $f_{D_{j}}(x) \ne f_{D_{l}}(x)$ when $j$ is an
interior node and $l$ is a boundary node.  As a result,
$D_{fix}+D_{j,i}$ is no longer symmetric if $j$ is a boundary
node. In fact, it is clear that the distribution of
$D_{fix}+D_{j,i}$ is a function of node $j$'s location near the
boundary.  Because of this additional asymmetry, let us assume for
a moment that the sequence of zero-crossings observed by boundary
node $i$ occur $\epsilon_{i}$ away from an integer value of $t$.
That is, if every node in the network, including the boundary
nodes, transmitted a sequence of pulses where each pulse was sent
according to (\ref{eq:pulse-connection-propDelay}), then boundary
node $i$ would observe the sequence of observations
\begin{equation} \label{eq:observation_form_boundary}
\alpha_{i}(\tau_{0}-m+l+\epsilon_{i}-\bar{\Delta}_{i})+\Psi_{i},
\end{equation}
where $l=0,1,\ldots,m-1$ and $\epsilon_{i}$ is known.

This boundary node $i$ could then use the time synchronization
estimator given by (\ref{eq:timesync-estimator-propDelay}) but
where the matrix $\bar{{\mathbf H}}$ is now replaced with
$\bar{{\mathbf H}}_{i}$
\begin{displaymath}
\bar{{\mathbf H}}_{i} = \left[ \begin{array}{ccccc}
1 & 1 & 1 & \ldots & 1\\
\epsilon_{i} & 1+\epsilon_{i} & 2+\epsilon_{i} & \ldots &
m-1+\epsilon_{i}
\end{array} \right]^T.
\end{displaymath}
Thus, for this boundary node $i$ we have
\begin{equation} \label{eq:timesync-estimator-propDelay-boundary}
V_{n,i}^{c_{i}}({\mathbf Y},\epsilon_{i}) = {\mathbf
C}(\bar{\mathbf{H}}_{i}^{T}\bar{\mathbf{H}}_{i})^{-1}\bar{\mathbf{H}}_{i}^{T}{\mathbf
Y},
\end{equation}
In this case, however, the variance of the time synchronization
estimator depends on $\epsilon_{i}$
\begin{equation} \label{eq:timesync-estimator-variance-boundary}
\textrm{Var}_{\theta}(V_{n,i}^{c_{i}}({\mathbf Y},\epsilon_{i})) =
\sigma^{2}\bigg(\frac{2(2m+1)}{m(m-1)}+\frac{12\epsilon_{i}(\epsilon_{i}-1-m)}{(m-1)m(m+1)}\bigg).
\end{equation}
The fact that the variance depends on $\epsilon_{i}$ is the root
of the problem.  The pulse-connection function
\begin{equation} \label{eq:pulse-connection-propDelay-boundary}
X_{n,i}^{c_{i}}({\mathbf Y})=V_{n,i}^{c_{i}}({\mathbf
Y},\epsilon_{i})+\hat{\alpha_{i}}D_{fix},
\end{equation}
is \emph{not} the same as that given by
(\ref{eq:pulse-connection-propDelay}).

To correct for this, we can make the strong assumption that each
boundary node $i$ knows is own $\alpha_{i}$.  We address the
reasoning behind this assumption in
Section~\ref{sec:propDelay_assumption}.  If we use this
assumption, then each boundary node $i$ can get an observation
sequence of the form (\ref{eq:observation_form}) simply by adding
$\alpha_{i}(\epsilon-\epsilon_{i})$ to each of the $m$
observations of the form given in
(\ref{eq:observation_form_boundary}), where we assume that node
$i$ knows both $\epsilon$ and $\epsilon_{i}$. With such an
observation sequence, boundary node $i$ will have the time
synchronization estimator (\ref{eq:timesync-estimator-propDelay})
and, more importantly, the pulse-connection function
(\ref{eq:pulse-connection-propDelay}). Thus, maintaining time
synchronization for the case of propagation delay would be
possible.

What we have then is that boundary node synchronization would
require only the boundary nodes to know their $\alpha_{i}$
parameters.  With this strong assumption only for the boundary
nodes, the network is effectively synchronized.  Even though the
boundary nodes do not see the same zero-crossing as the interior
nodes, they can calculate this time and thus have all the required
synchronization information.

\subsection{The Boundary Node Assumption}
\label{sec:propDelay_assumption}

The assumption that each boundary node $i$ knows $\alpha_{i}$ is a
strong assumption.  Even though the fraction of nodes that are
boundary nodes is small for multi-hop networks requiring many hops
to send information across the network, we believe that the
assumption is still very artificial. There are two reasons that we
make the assumption for the presentation of results on time
synchronization with propagation delay.

First, the assumption allows us to give an elegant presentation of
the main concept of this paper which is to use high node density
to average out errors in the network.  Throughout this work we
have used high node density to average out inherent errors present
in the nodes.  We were able to average out random timing jitter
that is present in each node and provide the network with a
sequence of zero-crossings that can serve as synchronization
events.  We then applied this technique to averaging out the
errors introduced by time delay.  To this end we were partially
successful in that the interior nodes can average out these errors
assuming the boundary nodes have additional information. But this
is of interest since the goal of this paper is to understand the
theory of spatial averaging for synchronization and discover its
fundamental advantages and limitations.

Second, the problem encountered at the boundaries is one that
opens up an entirely new area of study which is the target of our
future work. The issue that we encounter is that the waveform seen
by some nodes in the network will have a zero-crossing that is
shifted from the ideal location. This implies that different nodes
will observe different zero-crossings.  Furthermore, these
zero-crossings will now evolve in time since we do not have the
same observations over the entire network. This problem is similar
to what we encounter if we consider finite sized networks. For
finite $N$, the zero-crossing location will be random and thus
introduce another source of error. As well, different nodes will
see different zero-crossing locations. Therefore, we will turn our
attention to the case of finite $N$ and develop a different set of
tools that will be needed to understand what types of
synchronization are achievable under the situation where
zero-crossing locations evolve in time.  Using this understanding,
we hope to return to the issue of propagation delay in
asymptotically dense networks and characterize the behavior of the
network.

\section{Conclusions}
\label{sec:conclusion}

To conclude, we revisit the scalability issue under the light of
work developed in this paper.

\subsection{The Scalability Problem Revisited}

In the Introduction (Section~\ref{sec:estimate-params}), we
mentioned that most existing proposals for time synchronization
suffer from an inherent scalability problem.  The problem with
those existing proposals lies in the fact that synchronization
errors accumulate: if node 2 can synchronize to node 1 with some
small error, and node 3 can synchronize to node 2 with the same
small error, these errors accumulate, and the synchronization of
node 3 to node 1 is worse.  Therefore, synchronization error
increases with the number of hops in the network, and this problem
is especially apparent in the regime of high densities.  To make
these ideas precise, we first determine the maximum number of hops
over which synchronization information must travel and then study
how the error in a generic pairwise synchronization mechanism
depends on this number of hops.

\subsubsection{An Estimate of the Maximum Number of Hops}

To obtain an estimate for the maximum number of hops $\ell_N$ in a
network in the regime of high densities (fixed area,
$N\to\infty$), we approximate the transmission range of a node by
the minimum required transmission distance, $d_{N}$, to maintain a
fully connected network with high probability.
From~\cite{GuptaK:98}, we have that for $N$ nodes uniformly
distributed over a $[0,1]\times [0,1]$ square, the graph is
connected with probability-1 as $N\to\infty$ if and only if each
node's transmission distance $d_{N}$ is such that
\begin{displaymath}
\pi d_{N}^{2} = \frac{\log N + \epsilon_{N}}{N},
\end{displaymath}
for some $\epsilon_{N}\to\infty$.  Let us, therefore, approximate
$d_{N}$ as
\begin{displaymath}
d_{N} \approx \sqrt{\frac{1}{\pi}\frac{\log N}{N}}.
\end{displaymath}
Thus, $\ell_N = 1/d_N = O\big(\sqrt{\frac{N}{\log N}}\big)$, and thus
$\ell_N\to\infty$ as $N\to\infty$.

\subsubsection{Synchronization Error Over Multiple Hops}

Now, we assume there are $\ell_{N}$ nodes arranged in a linear
ordering, numbered $1$ to $\ell_{N}$.  To synchronize, each node
$i$ forms an estimate of its own $\alpha_{i}$, based on $m$ pulses
transmitted from node $i-1$. As before, node $1$ will have the
reference clock $c_{1}(t)=t$.

Node $1$ starts by sending $m$ pulses at times $\tau_{1}+l$ for
$l=0,1,\ldots,m-1$.  As a result, node $2$ will get a vector of
observations ${\mathbf Y}_{2}$, where ${\mathbf Y}_{2}[1] =
\alpha_{2}(\tau_{1}-\bar{\Delta}_{2})+\Psi_{2}$ and the $(l+1)$th
element of ${\mathbf Y}_{2}$ is ${\mathbf
Y}_{2}[l+1]=\alpha_{2}(\tau_{1}-\bar{\Delta}_{2})+l\alpha_{2}+\Psi_{2}$.
This is similar to the situation we had in (\ref{eq:simple-obs})
and we can therefore estimate $\alpha_{2}$ using
\begin{displaymath}
\hat{\alpha}_{2} = \bar{{\mathbf
C}}(\mathbf{H}^{T}\mathbf{H})^{-1}\mathbf{H}^{T}{\mathbf Y}_{2},
\end{displaymath}
where $\bar{{\mathbf C}} = [0\quad 1]$.  We find that
$\hat{\alpha}_{2} \sim {\mathcal N}(\alpha_{2},
12\sigma^{2}/((m-1)m(m+1)))$.

Node $2$ will now transmit $m$ pulses at times, in terms of
$c_{2}$, $\bar{\tau}_{2}+l\hat{\alpha}_{2}$, for
$l=0,1,\ldots,m-1$. Note that $\hat{\alpha}_{2}$ is now a fixed
value since node $2$ has estimated $\alpha_{2}$.  In terms of
$c_{1}$, these pulses occur at
\begin{displaymath}
(\bar{\tau}_{2}+l\hat{\alpha}_{2})^{c_{1}} =
\frac{\bar{\tau}_{2}+l\hat{\alpha}_{2}-\Psi_{2}}{\alpha_{2}}+\bar{\Delta}_{2}
= \tau_{2} + l\frac{\hat{\alpha}_{2}}{\alpha_{2}}-
\frac{\Psi_{2}}{\alpha_{2}},
\end{displaymath}
for $l=0,1,\ldots,m-1$, where $\tau_{2}=
(\bar{\tau}_{2}/\alpha_{2})+\bar{\Delta}_{2}$.  Thus, if we
translate these times into the time scale of $c_{3}$, we will have
the vector of observations, ${\mathbf Y}_{3}$, made by node $3$.
We find that the $(l+1)$th element of ${\mathbf Y}_{3}$ is
\begin{displaymath}
{\mathbf Y}_{3}[l+1] = \alpha_{3}((\tau_{2} +
l\frac{\hat{\alpha}_{2}}{\alpha_{2}}-
\frac{\Psi_{2}}{\alpha_{2}})-\bar{\Delta}_{3})+\Psi_{3} \sim
{\mathcal
N}\bigg(\alpha_{3}(\tau_{2}-\bar{\Delta}_{3})+l\alpha_{3}\frac{\hat{\alpha}_{2}}{\alpha_{2}},
\sigma^{2}\big(\frac{\alpha_{3}^{2}}{\alpha_{2}^{2}}+1\big)\bigg).
\end{displaymath}
This vector of observations is of the form
\begin{equation}
{\mathbf Y}_{3} = {\mathbf H}\bar{{\mathbf \theta}} +
\bar{{\mathbf W}},
\end{equation}
where
\begin{displaymath}
\bar{{\mathbf \theta}} = \left[ \begin{array}{c}
\bar{\theta}_{1}\\
\bar{\theta}_{2}
\end{array} \right] =
\left[ \begin{array}{c}
\alpha_{3}(\tau_{2}-\bar{\Delta}_{3})\\
\alpha_{3}\frac{\hat{\alpha}_{2}}{\alpha_{2}}
\end{array} \right]
\end{displaymath}
with
\begin{displaymath}
{\mathbf H} = \left[ \begin{array}{ccccc}
1 & 1 & 1 & \ldots & 1\\
0 & 1 & 2 & \ldots & m-1
\end{array} \right]^T
\end{displaymath}
and ${\mathbf W} = [W_{1}\dots W_{m}]^{T}$. ${\mathbf W} \sim
{\mathcal N}(0,\Sigma)$ with $\Sigma =
\sigma^{2}\big(\frac{\alpha_{3}^{2}}{\alpha_{2}^{2}}+1\big){\mathbf
I}$.

With this vector of observations, we can use the estimator
\begin{displaymath}
\hat{\alpha}_{3} = \bar{{\mathbf
C}}(\mathbf{H}^{T}\mathbf{H})^{-1}\mathbf{H}^{T}{\mathbf Y}_{3},
\end{displaymath}
where $\bar{{\mathbf C}} = [0\quad 1]$. We find that
\begin{displaymath}
\hat{\alpha}_{3} \sim {\mathcal
N}\bigg(\alpha_{3}\frac{\hat{\alpha}_{2}}{\alpha_{2}},
\frac{12\sigma^{2}}{((m-1)m(m+1))}\big(\frac{\alpha_{3}^{2}}{\alpha_{2}^{2}}+1\big)\bigg).
\end{displaymath}
If we continue this reasoning, we find that
\begin{displaymath}
\hat{\alpha}_{\ell_{N}} \sim {\mathcal
N}\bigg(\alpha_{\ell_{N}}\frac{\hat{\alpha}_{\ell_{N}-1}}{\alpha_{\ell_{N}-1}},
\frac{12\sigma^{2}}{((m-1)m(m+1))}\big(\frac{\alpha_{\ell_{N}}^{2}}{\alpha_{\ell_{N}-1}^{2}}+1\big)\bigg)
\end{displaymath}
will be the estimate of node $\ell_{N}$.

From the above analysis, we see that each node $i$'s estimate
suffers from jitter variance of the same form.  However, there is
an accumulation of error because node $i$'s estimate has a mean
that is dependent on node $i-1$'s estimate.  As a result, if node
$i-1$ has some small error, then that error will propagate to the
estimate of node $i$.  A good way to see this is if we consider
the special case where $\alpha_{2} = \alpha_{3} = \ldots
\alpha_{\ell_{N}} = 1$.  This is the case where the clock
frequencies are the same, but nodes do not know this.  In this
case, we find that node $\ell_{N}$'s estimate can be written as
\begin{displaymath}
\hat{\alpha}_{\ell_{N}} = \hat{\alpha}_{2} +
\sum_{i=3}^{\ell_{N}}W_{i}, \qquad \ell_{N}\geq 2
\end{displaymath}
where $W_{i}\sim{\mathcal N}(0,24\sigma^{2}/((m-1)m(m+1)))$. This
is intuitively obvious because node $i$'s estimate
$\hat{\alpha}_{i}$ will be the mean of the Gaussian random
variable $\hat{\alpha}_{i+1}$.  Therefore, it is obvious that the
error variance grows linearly with the number of hops.  In fact,
this behavior is observed in experimental work. With Reference
Broadcast Synchronization (RBS), from~\cite{ElsonGE:02} the
authors find that the synchronization error variance of an
$\ell_{N}$ hop path is approximately $\sigma^{2}\ell_{N}$, where
$\sigma^{2}$ is the one hop error variance. Therefore, we have
that the synchronization error between our two nodes will grow
linearly as $\ell_{N}=1/d_{N}$, which is strictly monotonically
increasing. As a result, as $N\to\infty$, we have that
synchronization error will grow unbounded.

This scalability problem, however, can potentially be avoided
using cooperative time synchronization as $N\to\infty$.  This is
because in the limit of infinite density, the cooperative time
synchronization technique allows every node in the network to see
a set of identical equispaced zero-crossings.  As a result, in
steady-state the synchronization error does not grow across the
network. This comes about by using the high node density to
average out random timing errors.  Thus, we find that cooperative
time synchronization has very favorable scalability properties in
the limit as $N\to\infty$.

\subsection{Network Density and Synchronization Performance Trade-Off}

The cooperative synchronization technique described in this paper
provides us deterministic parameters that we can use for time
synchronization in the limit as node density grows unbounded. In
fact, as the node density grows, the observations that can be used
for synchronization improve.  This means that our cooperative
synchronization technique provides an effective trade-off between
network density and synchronization performance. Such a trade-off
has not existed before and will provide network designers an
additional dimension over which to improve network synchronization
performance.

The fundamental idea behind cooperative time synchronization is
that by using spatial averaging, the errors inherent in each node
can be averaged out.  By using observations that are an ``average"
of the information from a large number of surrounding nodes,
synchronization performance can be improved due to the higher
quality observations.

\begin{figure}[!h]
\centerline{\psfig{file=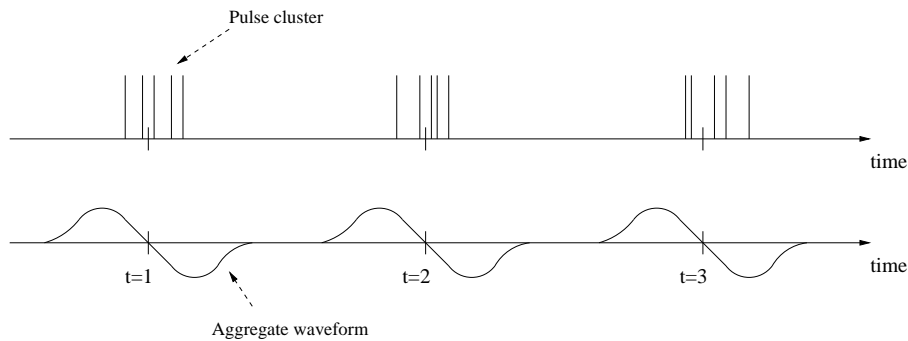,width=12cm}}
\vspace{-5mm} \caption[The connection with pulse-coupled
oscillators.]{\small Clusters of ultra wideband pulses can be used
for cooperative time synchronization.  In the the top figure, we
illustrate the clusters of pulses around integer values of $t$. As
the number of nodes increase, the sample mean will converge to the
integer value of the reference time.  This idea is parallel to the
use of zero-crossings shown in the bottom figure.}
\label{fig:pulsetraincluster}
\end{figure}

From this point of view, it is clear that the particular technique
described in this paper is but one example of using spatial
averaging to improve synchronization.  Other techniques can also
be developed using spatial averaging.  For example, nodes may not
necessarily have to send odd-shaped pulses and use zero-crossing
observations. Even though this setup takes advantage of the
superposition of pulses, it has its drawbacks.  To keep the
signals in phase, the jitter variance will limit the maximum
frequency at which signals can be sent.  Instead, nodes may
transmit ultra wideband pulses.  If the nodes surrounding a
particular node $j$ each transmit an impulse at their estimate of
an integer value of $t$, then due to timing errors in the
surrounding nodes, node $j$ will see a cluster of pulse arrivals
around this integer value of $t$.  Node $j$ can then take the
sample mean of this cluster of pulses and use that as an
observation, just like we used the zero-crossing as an observation
in this paper.  This idea is illustrated in
Fig.~\ref{fig:pulsetraincluster}. Such a technique based on ultra
wideband pulses will also provide similar scalability properties.
As a result, cooperative time synchronization really describes a
class of techniques that can take advantage of spatial averaging
to improve synchronization performance.

\subsection{Future Work}

With the goal of developing practical cooperative synchronization
mechanisms, two keys areas of interest are cooperative
synchronization in finite-sized networks and algorithm
development.  First, the analysis of performance for finite-sized
networks is very important. Determining when the asymptotic
properties presented in this work are good predictors of
performance in networks that may be large but still finite in size
is important in terms of bridging the gap between our proposed
ideas and practical systems. Preliminary, simulation-based work
along these lines can be found in~\cite{HuS:03b}.  Second,
developing practical techniques for cooperative time
synchronization is essential for implementing spatial averaging in
real networks.  Along these lines, one area of interest is
determining what types of pulses should be used, i.e. odd-shaped
pulses or ultra wideband pulses.

Furthermore, the ideas in this paper suggest a few other areas of
interest for future work.  One is the issue of distributed
modulation methods.  If we have the ability to generate an
aggregate waveform with equispaced zero-crossings, by controlling
the location of these crossings we can modulate information onto
this waveform and use it to communicate with a far receiver.
Preliminary work along these lines can be found in~\cite{HuS:03c}.
Another issue is to study how the idea of spatial averaging that
is so prevalent in this work contributes to synchronization that
is observed in nature.

\pagebreak
\appendix

{\bf Proof of Lemma~\ref{lemma:polarity_positive}.} \hspace{0.5cm}
To show~(\ref{eqn:cond3a}), we consider
\begin{eqnarray*}
E(A_{max}K_{i}p(\tau_{1}-\tau_{0}-T_{i})) &=& A_{max}E(K_{i})E(p(\tau_{1}-\tau_{0}-T_{i})) \\
&=& A_{max}E(K_{i})\int p(\tau_{1}-\tau_{0}-\psi)f_{T_{i}}(\psi)d\psi \\
&=& -A_{max}E(K_{i})\int
p(\psi-(\tau_{1}-\tau_{0}))f_{T_{i}}(\psi)d\psi
\end{eqnarray*}
Since $\tau_{1}<\tau_{0}$, we have that $\tau_{1}-\tau_{0}<0$
implying that $p(\psi)$ is shifted to the left and the
zero-crossing of $p(\psi)$ occurs at a negative value. $p(\psi)$
is odd about its zero-crossing and $f_{T_{i}}(\psi)$ is symmetric
about zero and strictly monotonically increasing on $(-\infty,0]$
for all positive finite variance values.  Thus, it is clear that
$\int p(\psi-(\tau_{1}-\tau_{0}))f_{T_{i}}(\psi)d\psi <0$ which
makes $E(A_{max}K_{i}p(\tau_{1}-\tau_{0}-T_{i}))>0$.

Now, the expectation will vary with the variance of $T_{i}$ and
the variance will range from a positive upper bound of
$\bar{\sigma}^{2}/\alpha_{low}^{2}<B$ to a positive lower bound of
$\bar{\sigma}^{2}/\alpha_{up}^{2}$, where recall that
$\bar{\sigma}^{2}$ is a value determined by our choice of the
pulse connection function. If we consider $\int
p(\psi-(\tau_{1}-\tau_{0}))f_{T_{i}}(\psi)d\psi$ to be a function
of the variance of $T_{i}$, then we see that it is bounded and
continuous on the compact domain
$[\bar{\sigma}^{2}/\alpha_{up}^{2},
\bar{\sigma}^{2}/\alpha_{low}^{2}]$. Since we showed in the
previous paragraph that
$E(A_{max}K_{i}p(\tau_{1}-\tau_{0}-T_{i}))>0$ whenever $T_{i}$ has
a nonzero finite variance, clearly
$E(A_{max}K_{i}p(\tau_{1}-\tau_{0}-T_{i}))>0$ when
$\textrm{Var}(T_{i})\in [\bar{\sigma}^{2}/\alpha_{up}^{2},
\bar{\sigma}^{2}/\alpha_{low}^{2}]$. Thus, it is clear that
$\gamma_{1}$ and $\gamma_{2}$ exist and~(\ref{eqn:cond3a}) is
shown.

To show~(\ref{eqn:cond3b}), we consider
\begin{eqnarray*}
\textrm{Var}(A_{max}K_{i}p(\tau_{1}-\tau_{0}-T_{i}))&=& E(A_{max}^{2}K^{2}_{i}p^{2}(\tau_{1}-\tau_{0}-T_{i}))-E^{2}(A_{max}K_{i}p(\tau_{1}-\tau_{0}-T_{i})) \\
&\leq& A_{max}^{2}E(K^{2}_{i})E(p^{2}(\tau_{1}-\tau_{0}-T_{i}))\\
&\leq& A_{max}^{2}E(K^{2}_{i})\\
&\leq& A_{max}^{2}
\end{eqnarray*}
where the second to last inequality follows from the fact that
$E(p^{2}(\tau_{1}-\tau_{0}-T_{i}))$ is upper bounded by $1$. The
last inequality follows since $E(K^{2}_{i})\leq 1$ by the fact
that $0\leq K_{i}\leq 1$. Thus, we have shown~(\ref{eqn:cond3b}).

Next we define $S_{n} = \bar{M}_{1}(\tau_{1}) + \dots +
\bar{M}_{n}(\tau_{1})$ and $m_{n} = E(S_{n}) =
\mu_{1}+\ldots+\mu_{n}$. From~\cite{Feller:50} we have the
following theorem
\begin{theorem} \label{theorem:Feller}
The convergence of the series
\begin{displaymath}
\sum \frac{\sigma_{i}^{2}}{i^{2}}
\end{displaymath}
implies that the strong law of large numbers will apply to the
sequence of independent random variables $\bar{M}_{i}(\tau_{1})$.
That is, again from~\cite{Feller:50}, for every pair $\epsilon>0$,
$\delta>0$, there corresponds an $N$ such that
\begin{displaymath}
\textrm{Pr}\bigg\{\frac{|S_{n}-m_{n}|}{n}<\epsilon;\quad
n=N,N+1,\ldots,N+r\bigg\}
> 1-\delta
\end{displaymath}
for all $r>0$. \qquad $\bigtriangleup$
\end{theorem}

We have shown~(\ref{eqn:cond3b}) so we have $\sigma_{i}^{2}<\gamma_{3}<\infty$.  Thus
\[
\lim_{N\to\infty} \sum_{i=1}^{N} \frac{\sigma_{i}^{2}}{i^{2}} \;\; \leq \;\;
\lim_{N\to\infty} \sum_{i=1}^{N} \frac{\gamma_{3}}{i^{2}} \;\; = \;\;
\gamma_{3}\frac{\pi^{2}}{6}.
\]
and we have convergence by the direct comparison test.  Therefore,
we can apply Theorem~\ref{theorem:Feller} and get that for any
pair $\epsilon>0$, $\delta>0$, we can find an $N$ such that
\begin{eqnarray} \label{eqn:slln}
\textrm{Pr}\bigg\{\bigg|\frac{S_{n}}{n}-\frac{m_{n}}{n}\bigg|<\epsilon;\quad
n=N,N+1,\ldots,N+r\bigg\} > 1-\delta
\end{eqnarray}
for all $r>0$.

By~(\ref{eqn:cond3a}) we have that $\gamma_{2}>\mu_{i} > \gamma_{1} > 0$.  Thus, we can clearly see that
\begin{displaymath}
\frac{m_{n}}{n} > \gamma_{1}.
\end{displaymath}
Furthermore, since we keep the function $f_{\alpha}(s)$ constant
as we increase the number of nodes in the network we get that
$m_{n}/n$ converges to a constant $\eta(\tau_{1})$ given by
\begin{eqnarray*}
\eta(\tau_{1}) &=&
A_{max}E(K_{i})\int_{\alpha_{low}}^{\alpha_{up}}
\int_{-\infty}^{\infty} p(\tau_1-\tau_0-\psi) f_{T}(\psi,s) d\psi
f_{\alpha}(s) ds \\
&=& \int_{\alpha_{low}}^{\alpha_{up}} E(\bar{M}_{i}(\tau_{1},s))
f_{\alpha}(s) ds.
\end{eqnarray*}
The above expression comes from the fact that since each
$\mu_{i}=E(\bar{M}_{i}(\tau_{1}))$ is a function of $\alpha_{i}$,
$m_{n}/n$ will converge to the average of the $\mu_{i}$ over
$f_{\alpha}(s)$, the function that characterizes the set of
$\alpha_{i}$'s. Therefore, given any $\epsilon$, we can find an
$N'$ such that
\begin{equation} \label{eq:means-converge}
\bigg|\frac{m_{n}}{n}-\eta(\tau_{1})\bigg|< \epsilon
\end{equation}
for all $n>N'$.  Note that since $(m_{n}/n)>\gamma_{1}$, we have
that $\eta(\tau_{1})\geq \gamma_{1}$. Since
\begin{displaymath}
\bigg|\frac{S_{n}}{n}-\eta(\tau_{1})\bigg| <
\bigg|\frac{S_{n}}{n}-\frac{m_{n}}{n}\bigg|+
\bigg|\frac{m_{n}}{n}-\eta(\tau_{1})\bigg|,
\end{displaymath}
using (\ref{eqn:slln}) and (\ref{eq:means-converge}) we have
\begin{eqnarray*}
\textrm{Pr}\bigg\{\bigg|\frac{S_{n}}{n}-\eta(\tau_{1})\bigg|<2\epsilon;\quad
n=N'',N''+1,\ldots,N''+r\bigg\} > 1-\delta.
\end{eqnarray*}
for all $r>0$, where $N''=\max\{N,N'\}$. Thus, we have
\begin{displaymath}
\lim_{N\to\infty} \frac{1}{N} \sum_{i=1}^{N} \bar{M}_{i}(\tau_{1})
=\eta(\tau_{1})
>0
\end{displaymath}
almost surely. This completes the proof of Lemma~\ref{lemma:polarity_positive}. \qquad
$\bigtriangleup$

{\bf Proof of Lemma~\ref{lemma:continuity1}.} \hspace{0.5cm}
First, we start by finding an analytical expression for
$|A_{\infty}(t)-A_{\infty}(t_{o})|$. From the proof of
Lemma~\ref{lemma:polarity_positive} we have that
\begin{displaymath}
A_{\infty}(t) = A_{max}E(K_{i})\int_{\alpha_{low}}^{\alpha_{up}}
\int_{-\infty}^{\infty} p(t-\tau_0-\psi) f_{T}(\psi,s) d\psi
f_{\alpha}(s) ds.
\end{displaymath}
Therefore, $|A_{\infty}(t)-A_{\infty}(t_{o})|$ can be written as
\begin{eqnarray*}
\lefteqn{|A_{\infty}(t)-A_{\infty}(t_{o})|} \\
& = &
|A_{max}E(K_{i})\int_{\alpha_{low}}^{\alpha_{up}}\int_{-\infty}^{\infty}
[p(t-\tau_{o}-\psi)-p(t_{o}-\tau_{o}-\psi)]
f_{T}(\psi,s) f_{\alpha}(s)d\psi ds| \\
& \leq & A_{max} \int_{\alpha_{low}}^{\alpha_{up}}\int_{-\infty}^{\infty} |p(t-\tau_{o}-\psi)-p(t_{o}-\tau_{o}-\psi)|f_{T}(\psi,s) f_{\alpha}(s)d\psi ds \\
& = &  A_{max}
\int_{\alpha_{low}}^{\alpha_{up}}\int_{-\tau_{nz}+t_{o}-\tau_{0}-|t-t_{o}|}^{\tau_{nz}+t_{o}-\tau_{0}+|t-t_{o}|}
|p(t-\tau_{o}-\psi)-p(t_{o}-\tau_{o}-\psi)|
f_{T}(\psi,s)f_{\alpha}(s) d\psi ds,
\end{eqnarray*}
where $E(K_{i})\leq 1$. The change in the limits of integration in
the last equality comes from the fact that
$p(t-\tau_{o}-\psi)-p(t_{o}-\tau_{o}-\psi)=0$ outside of $\psi \in
[-\tau_{nz}+t_{o}-\tau_{0}-|t-t_{o}|,\tau_{nz}+t_{o}-\tau_{0}+|t-t_{o}|]$.
This is the maximum interval over which
$p(t-\tau_{o}-\psi)-p(t_{o}-\tau_{o}-\psi)$ can be non-zero. There
is no need to take the absolute value of $f_{T}(\psi,s)$ and
$f_{\alpha}(s)$ since they are always non-negative.

Our second step is to bound the inner integral.  Before doing so,
we first show that the inside integral is in fact Riemann
integrable. For any given $t$ and $t_{o}$, the inside integral is
taken over a closed interval.  Over a closed interval, we know
from Strichartz~\cite{Strichartz:00} that any bounded function
that is continuous except at a finite number of points is Riemann
integrable.  Furthermore, also from~\cite{Strichartz:00} we know
that the sums and products of continuous functions are continuous.
As well, if a function is continuous then the absolute value of
that function is also continuous. $p(t)$ has at most $D=3$
locations at which it is discontinuous and over any open interval
not containing a discontinuity, $p(t)$ is uniformly continuous
since $q(t)$ is uniformly continuous. $f_{T}(\psi,s)$ has $D'=0$
discontinuities in $\psi$ for an given $s$ since it is Gaussian
for any $s$.  And since $s\in [\alpha_{low},\alpha_{up}]$,
$|f_{T}(\psi,s)|\leq G_{T}$ for all $\psi$ and $s$ ($G_{T}$
occurring when $\psi=0$ and $s=\alpha_{up}$). Thus, since $p(t)$
and $f_{T}(\psi,s)$ are continuous except at a finite number of
points, we see that for given $s$, $t$, and $t_{0}$
\begin{displaymath}
|p(t-\tau_{o}-\psi)-p(t_{o}-\tau_{o}-\psi)| f_{T}(\psi,s)
\end{displaymath}
is also continuous in $\psi$ except at a finite number of points
(at most $D'+2D$ points). This function is also bounded since the
product of two bounded functions is bounded.  As a result, we see
that the integral is Riemann integrable over any closed interval.

We now proceed to bound from above the value of this integral by
first bounding the maximum value of the integral assuming no
discontinuities and then introducing another term that bounds the
maximum area contributed by the discontinuities. If we ignore the
discontinuities and assume $p(t)$ is uniformly continuous, for any
$m_{1}>0$ there exists a $n>0$ such that
\begin{displaymath}
|t-t_{o}|<\frac{1}{n} \Rightarrow |p(t)-p(t_{o})|<\frac{1}{m_{1}},
\end{displaymath}
for all $t$ and $t_{o}$.  As a result,
$p(t-\tau_{o}-\psi)-p(t_{o}-\tau_{o}-\psi)$ can be made as small
as desired by choosing the proper $n$ thus giving us
$p(t-\tau_{o}-\psi)-p(t_{o}-\tau_{o}-\psi)<1/m_{1}$ for all $\psi$
for an appropriate choice of $n$.

Furthermore, we note that $|p(t-\tau_{o}-\psi)| f_{T}(\psi,s) \leq
G_{T}$ because $|p(t)|\leq 1$ and $|f_{T}(\psi,s)|\leq G_{T}$. The
maximum possible jump at a discontinuity in the function
$|p(t-\tau_{o}-\psi)-p(t_{o}-\tau_{o}-\psi)| f_{T}(\psi,s)$ is
thus $2G_{T}$ and for any $|t-t_{o}|$, the maximum area
contributed by each discontinuity is $2G_{T}|t-t_{o}|$.  As a
result, for all $D'+2D$ discontinuities, the maximum area
contribution will be no more than $2G_{T}|t-t_{o}|(D'+2D)$.

We can, therefore,  bound the inner integral as
\begin{eqnarray*}
\lefteqn{\int_{-\tau_{nz}+t_{o}-\tau_{0}-|t-t_{o}|}^{\tau_{nz}+t_{o}-\tau_{0}+|t-t_{o}|}
|p(t-\tau_{o}-\psi)-p(t_{o}-\tau_{o}-\psi)| f_{T}(\psi,s) d\psi}
\\  & \leq &
\int_{-\tau_{nz}+t_{o}-\tau_{0}-|t-t_{o}|}^{\tau_{nz}+t_{o}-\tau_{0}+|t-t_{o}|} \frac{G_{T}}{m_{1}} d\psi+ 2G_{T}|t-t_{o}|(D'+2D) \\
& = & \frac{G_{T}}{m_{1}}(2\tau_{nz}+2|t-t_{o}|)+ 2G_{T}|t-t_{o}|(D'+2D) \\
& = & 2\frac{G_{T}}{m_{1}}\tau_{nz}+2\frac{G_{T}}{m_{1}}|t-t_{o}|
+ 2G_{T}|t-t_{o}|(D'+2D),
\end{eqnarray*}
where $|t-t_{o}|<1/n$.

What we have is that if $|t-t_{0}|<1/n$ then
\begin{eqnarray*}
\lefteqn{|A_{\infty}(t)-A_{\infty}(t_{o})|} \\
& \leq & A_{max}
\int_{\alpha_{low}}^{\alpha_{up}}\bigg(2\frac{G_{T}}
{m_{1}}\tau_{nz}+2\frac{G_{T}}{m_{1}}|t-t_{o}|
+ 2G_{T}|t-t_{o}|(D'+2D)\bigg) f_{\alpha}(s) ds \\
& \leq & A_{max}G_{\alpha}(\alpha_{up}-\alpha_{low})
\bigg(2\frac{G_{T}}{m_{1}}\tau_{nz}+2\frac{G_{T}}{m_{1}}|t-t_{o}|
+ 2G_{T}|t-t_{o}|(D'+2D)\bigg)
\end{eqnarray*}
since $|f_{\alpha}(s)|<G_{\alpha}$ (defined in
Section~\ref{sec:systemParameters}). We define $\bar{A}$ as
\begin{displaymath}
\bar{A}=A_{max}G_{\alpha}(\alpha_{up}-\alpha_{low}).
\end{displaymath}

Now, for the third step of our proof we make
\begin{eqnarray*}
\lefteqn{|A_{\infty}(t)-A_{\infty}(t_{o})|}\\
& \leq & \bar{A}\bigg(2\frac{G_{T}}{m_{1}}\tau_{nz}+2\frac{G_{T}}{m_{1}}|t-t_{o}|+ 2G_{T}|t-t_{o}|(D'+2D)\bigg)\\
&<& \frac{1}{m},
\end{eqnarray*}
for any choice of $m>0$. We do this by making each of the three
terms less than $1/(3m)$.

For the first term we want
\begin{displaymath}
\frac{2\bar{A}G_{T}\tau_{nz}}{m_{1}} < \frac{1}{3m}.
\end{displaymath}
We solve and get
\begin{displaymath}
m_{1} > 6m\bar{A}G_{T}\tau_{nz}.
\end{displaymath}
Since for any value of $m_{1}>0$ we can find an $n>0$, this
condition can be satisfied.

For the third term we want
\begin{displaymath}
2\bar{A}G_{T}(D'+2D)|t-t_{o}| < \frac{1}{3m}.
\end{displaymath}
This gives us
\begin{displaymath}
|t-t_{o}|< \frac{1}{6\bar{A}G_{T}(D'+2D)m}.
\end{displaymath}
Since the only requirement is $|t-t_{o}|<1/n$ for $n$ chosen by
any given $m_{1}>0$, we can always choose $|t-t_{o}|$ as small as
desired.  Thus, this condition can be satisfied.

With the second term we want the condition
\begin{displaymath}
\frac{2\bar{A}G_{T}}{m_{1}}|t-t_{o}| < \frac{1}{3m}
\end{displaymath}
which means that
\begin{displaymath}
\frac{|t-t_{o}|}{m_{1}} < \frac{1}{6m\bar{A}G_{T}}.
\end{displaymath}
Again, this condition can be satisfied since we can choose $m_{1}$
as large as we want and $|t-t_{o}|$ as small as we want as long as
$|t-t_{o}|<1/n$ for a given $m_{1}$.

Thus, for any $m>0$, we first choose
$m_{1}>6m\bar{A}G_{T}\tau_{nz}$. Then, we find an $n'>0$ such that
$|t-t_{o}|<1/n'$ implies that $|p(t)-p(t_{o})|<1/m_{1}$ for all
$t$ and $t_{o}$ if we remove the discontinuities in $p(t)$. Then,
if necessary, $n'$ is increased to $n$ so that $|t-t_{o}|<1/n$
implies that $|t-t_{o}|< 1/(6\bar{A}G_{T}(D'+2D)m)$ and
$|t-t_{o}|/m_{1} < 1/(6m\bar{A}G_{T})$. If no increase is
necessary, then $n=n'$. With this choice of $n>0$,
$|A_{\infty}(t)-A_{\infty}(t_{o})|<1/m$.  As a result, for any
$m$, we can find an $n$ such that $|t-t_{o}|<1/n$ implies that
$|A_{\infty}(t)-A_{\infty}(t_{o})| < 1/m$.  Thus, $A_{\infty}(t)$
is continuous.

This completes the proof for Lemma \ref{lemma:continuity1}.
$\bigtriangleup$

{\bf Proof of Theorem~\ref{theorem:main-delay}.} \hspace{0.5cm}
Let us start by writing
(\ref{eq:timesync-aggwaveform-withDelayAndFix}) as
\begin{displaymath}
A^{c_{1}}_{j,N}(t)\;\; = \;\;\sum_{i=1}^{N}
\frac{A_{max}K_{fix}K_{j,i}}{N}
p(t-\tau_{o}-T_{i}-D_{fix}-D_{j,i}) \;\; = \;\; \sum_{i=1}^{N}
\frac{1}{N} \tilde{M}_{i}(t,s),
\end{displaymath}
where $\tilde{M}_{i}(t,s)\stackrel{\Delta}{=}A_{max}K_{fix}K_{j,i}
p(t-\tau_{o}-T_{i}-D_{fix}-D_{j,i})$. Recall that the dependence
on $s$ comes from the fact that the density of $T_{i}$ is a
function of $\alpha_{i}$ which is characterized by
$f_{\alpha}(s)$. This notation is analogous to the notation used
in Section~\ref{sec:pulseProperties}. Following the steps in the
proof of Lemma~\ref{lemma:polarity_positive}, we can quickly show
that the limiting aggregate waveform at node $j$ will take on the
form
\begin{eqnarray} \label{eq:limitingWave-delayProof}
\eta(t) =
\int_{\alpha_{low}}^{\alpha_{up}}E(\tilde{M}_{i}(t,s))f_{\alpha}(s)ds,
\end{eqnarray}
where
\begin{eqnarray*}
\lefteqn{E(\tilde{M}_{i}(t,s)) } \\
 & & = A_{max}\int_{-\infty}^{\infty}
\int_{-\infty}^{0}
\int_{0}^{\infty}g(-y)g(x)p(t-\tau_{0}-\psi-y-x)f_{D_{j}}(x)
f_{D_{fix}}(y) f_{T}(\psi,s)dx dy d\psi,
\end{eqnarray*}
with $g(\cdot)=K(\delta^{-1}(\cdot))$. Therefore, we can prove
Theorem~\ref{theorem:main-delay} in two steps:\begin{itemize}
\item To show that $\eta(t)$ is odd about $\tau_{0}$, we need to
show that $E(\tilde{M}_{i}(t,s))$ is odd in $t$ about $\tau_{0}$,
i.e.
$E(\tilde{M}_{i}(\tau_{0}+\xi,s))=-E(\tilde{M}_{i}(\tau_{0}-\xi,s))$
for $\xi\geq0$. \item To show a zero-crossing at $\tau_{0}$, show
that $E(\tilde{M}_{i}(\tau_{0},s))=0$.
\end{itemize}
These two steps come directly from the form of $\eta(t)$ in
(\ref{eq:limitingWave-delayProof}).

We first show that
$E(\tilde{M}_{i}(\tau_{0}+\xi,s))=-E(\tilde{M}_{i}(\tau_{0}-\xi,s))$
for $\xi\geq 0$. Using the fact that
$K_{fix}=K(\delta^{-1}(-D_{fix}))=g(-D_{fix})$ and
$K_{j,i}=g(D_{j,i})$, we have the following:
\begin{eqnarray*}
\lefteqn{E(\tilde{M}_{i}(\tau_{0}+\xi,s))} \\ &=&
E\big(A_{max}g(-D_{fix})
g(D_{j,i}) p(\xi-[T_{i}+D_{fix}+D_{j,i}])\big) \\
&\stackrel{\small{(a)}}{=}& -E\big(A_{max}g(-D_{fix})
g(D_{j,i}) p(-\xi+[T_{i}+D_{fix}+D_{j,i}])\big) \\
&=& -A_{max}\int_{-\infty}^{\infty} \int_{-\infty}^{0}
\int_{0}^{\infty}g(-y)g(x)p(-\xi+[\psi+y+x])f_{D_{j}}(x)
f_{D_{fix}}(y) f_{T}(\psi,s)dx dy d\psi \\
&\stackrel{\small{(b)}}{=}& A_{max}\int_{\infty}^{-\infty}
\int_{\infty}^{0}
\int_{0}^{-\infty}g(z)g(-u)p(-\xi-[w+z+u])f_{D_{j}}(-u)
f_{D_{fix}}(-z) f_{T}(-w,s)du dz dw \\
&\stackrel{\small{(c)}}{=}& -A_{max}\int_{-\infty}^{\infty}
\int_{-\infty}^{0}
\int_{0}^{\infty}g(-u)g(z)p(-\xi-[w+u+z])f_{D_{j}}(z)
f_{D_{fix}}(u) f_{T}(w,s)dz du dw \\
&=& -E\big(A_{max}g(-D_{fix})
g(D_{j,i}) p(-\xi-[T_{i}+D_{fix}+D_{j,i}])\big) \\
&=& -E(\tilde{M}_{i}(\tau_{0}-\xi,s)),
\end{eqnarray*}
where $(a)$ follows because $p(t)=-p(-t)$ and at $(b)$ we did a
change of variables with $u=-x$, $w=-\psi$, and $z=-y$.  $(c)$
follows from $f_{T}(x,s)=f_{T}(-x,s)$ and
$f_{D_{j}}(x)=f_{D_{fix}}(-x)$.  We thus have
$E(\tilde{M}_{i}(\tau_{0}+\xi,s))=-E(\tilde{M}_{i}(\tau_{0}-\xi,s))$
for $\xi\geq0$.

$E(\tilde{M}_{i}(\tau_{0},s))=0$ can now be shown as follows.
Using the just proven fact that
$E(\tilde{M}_{i}(\tau_{0}+\xi,s))=-E(\tilde{M}_{i}(\tau_{0}-\xi,s))$
for $\xi\geq0$, setting $\xi=0$ gives us
$E(\tilde{M}_{i}(\tau_{0},s))=-E(\tilde{M}_{i}(\tau_{0},s))$. This
implies that $E(\tilde{M}_{i}(\tau_{0},s))=0$.

This completes the proof for Theorem~\ref{theorem:main-delay}.
$\bigtriangleup$


\pagebreak

\begin{biography}{An-swol Hu} was born in New York State and grew
up in California.  He received his B.S. in Electrical Engineering
from Stanford University in 2002.  Currently he is a Ph.D.
candidate in the School of Electrical and Computer Engineering at
Cornell University.  His research interests include applied
statistics and statistical signal processing, with applications to
sensor networks.
\end{biography}

\begin{biography}{Sergio D.\ Servetto} was born in Argentina, on
January 18, 1968.  He received a Licenciatura en Informatica from
Universidad Nacional de La Plata (UNLP, Argentina) in 1992, and
the M.Sc. degree in Electrical Engineering and the Ph.D. degree in
Computer Science from the University of Illinois at
Urbana-Champaign (UIUC), in 1996 and 1999.  Between 1999 and 2001,
he worked at the Ecole Polytechnique Federale de Lausanne (EPFL),
Lausanne, Switzerland. Since Fall 2001, he has been an Assistant
Professor in the School of Electrical and Computer Engineering at
Cornell University, and a member of the field of Applied
Mathematics.  He was the recipient of the 1998 Ray Ozzie
Fellowship, given to ``outstanding graduate students in Computer
Science,'' and of the 1999 David J. Kuck Outstanding Thesis Award,
for the best doctoral dissertation of the year, both from the
Dept. of Computer Science at UIUC.  He is also the recipient of a
2003 NSF CAREER Award.  His research interests are centered around
information theoretic aspects of networked systems, with a current
emphasis on problems that arise in the context of large-scale
sensor networks.

\end{biography}

\end{document}